\documentclass[a4paper,12pt] {article}
\usepackage{graphics}
\usepackage{graphicx}
\usepackage{pstricks}
\usepackage[english]{babel}
\usepackage{amsmath}
\usepackage{amssymb}
\usepackage{float}
\usepackage[caption=false]{subfig}
\usepackage{multirow}
\usepackage{bm}
\usepackage{url}
\usepackage{placeins}
\usepackage{pdfpages}
\usepackage[abs]{overpic}
\usepackage{pict2e}

\definecolor{lgray}{gray}{0.85}
\definecolor{dgray}{RGB}{128,128,128}
\definecolor{lblue}{RGB}{0,191,191}
\definecolor{dblue}{RGB}{0,114,189}
\definecolor{orange}{RGB}{222,125,0}
\definecolor{dorange}{RGB}{217,83,25}
\definecolor{dred}{RGB}{217,83,25}
\definecolor{dgreen}{RGB}{119,172,48}
\definecolor{dyellow}{RGB}{255,215,0}

\begin{document}
\title{\Large\textbf{Application of control-based continuation to a nonlinear structure with harmonically coupled modes.}}
\author{\textbf{L.~Renson$^{1,}$\footnote{Corresponding author: l.renson@bristol.ac.uk} , A.D.~Shaw$^{2}$, D.A.W.~Barton$^{1}$, S.A.~Neild$^{1}$} \vspace{2mm}\\
\normalsize{$^{1}$Faculty of Engineering, University of Bristol, UK.}\\
\normalsize{$^{2}$College of Engineering, Swansea University, UK.}}
%\date{\normalsize{\today}}
\date{}
\maketitle

\begin{abstract}
This paper presents a systematic method for exploring the nonlinear dynamics of multi-degree-of-freedom (MDOF) physical experiments. To illustrate the power of this method, known as control-based continuation (CBC), it is applied to a nonlinear beam structure that exhibits a strong 3:1 modal coupling between its first two bending modes. CBC is able to extract a range of dynamical features, including an isola, directly from the experiment without recourse to model fitting or other indirect data-processing methods.\\
 
Previously, CBC has only been applied to (essentially) single-degree-of-freedom experiments; in this paper we show that the required feedback-control methods and path-following techniques can equally be applied to MDOF systems. A low-level broadband excitation is initially applied to the experiment to obtain the requisite information for controller design and, subsequently, the physical experiment is treated as a “black box” that is probed using CBC. The invasiveness of the controller used is analysed and experimental results are validated with open-loop measurements. Good agreement between open- and closed-loop results is achieved, though it is found that care needs to be taken in dealing with the presence of higher-harmonics in the force applied to the structure.\\ 

\end{abstract}

\section{Introduction}
As mechanical structures are increasingly designed to be lighter and more flexible than historically has been the case, nonlinear behaviour is increasingly being observed during experimental tests. Take, for example, the Airbus A400-M aircraft where ground vibration tests revealed the presence of nonlinearities affecting the aircraft resonant frequencies and mode shapes~\cite{Ahlquist}, or the qualification tests of the SmallSat satellite where nonlinear interactions between modes due to the presence of harmonics in the response were observed~\cite{Noel13,Renson14}. In~\cite{Ehrhardt17}, cracks in a circular perforated plate were thought to be caused by the change of stress distribution induced by similar modal interaction phenomena.\\

The systematic characterisation of the dynamic behaviour of multi-degree-of-freedom (MDOF) nonlinear structures is challenging and expensive in terms of experimental effort, particularly when compared with linear systems which can be characterised very efficiently with broadband modal analysis techniques. The presence, in nonlinear systems, of multiple coexisting steady-state responses to identical inputs requires repeated tests to capture the possible behaviours. Determining, precisely and with confidence, the regions where these multiple responses exist further multiplies the testing effort as experimental errors such as noise can lead to untimely transitions between these responses even before the region boundaries are reached.\\

Nonlinear systems can also exhibit a wide variety of other complex phenomena not seen in linear systems such as modal interactions~\cite{VakakisTET}, quasi-periodic oscillations~\cite{NayfehBook} and isola~\cite{Kuether15}. The latter correspond to branches of stable responses that are isolated from the main peak of the nonlinear frequency response of the system. Isola can arise through a number of mechanisms such as sub- and super-harmonic resonances on simple single-degree-of-freedom (SDOF) nonlinear systems~\cite{NayfehBook,Bureau14,Elmegard14} as well as through modal interactions. As such, isola have been observed across a wide range of nonlinear mechanical systems, for instance, suspended cables~\cite{Rega05}, pedestrian bridges~\cite{Lenci09}, vibration absorbers~\cite{Alexander09,Detroux15b} and a satellite structure~\cite{Detroux15}. The reader is referred to~\cite{Habib18} for a more detailed review on this topic.\\

The aforementioned challenges associated with nonlinear MDOF structures mean that a robust systematic method for experimental testing is highly desirable. This paper explores the use of control-based continuation (CBC) for this purpose. CBC exploits feedback control to provide a number of advantages compared with classical open-loop approaches to experimental testing. The controller maintains the experiment around a prescribed operating point, thus avoiding untimely transitions between coexisting behaviours. Unstable responses of the underlying uncontrolled system can also be made stable, and hence observable. Observing both stable and unstable responses provides a more complete picture of the system's dynamics and how different responses are interconnected, which facilitates the interpretation of the dynamics.\\

With CBC the controller is made non-invasive by an iterative scheme such that the positions in parameter space of the responses of the closed-loop experiment are identical to the open-loop (uncontrolled) experiment. Through the application of methods from the dynamical systems community, the evolution of the responses can be tracked as parameters (for instance, forcing frequency/amplitude) are varied and nonlinear dynamic features (such as nonlinear frequency responses, nonlinear normal modes, or bifurcation curves) can be extracted directly from the experiment without recourse to indirect methods that rely on post-processing of the data. These features can in turn be exploited for the development and validation of mathematical models, for which a number of techniques already exist~\cite{Peter15,Hill15,Song18}.\\

CBC was originally presented by Sieber and Krauskopf~\cite{Sieber08}. Using simple proportional-plus-derivative controllers, CBC was used to trace out the periodic oscillations of a parametrically-excited pendulum~\cite{Sieber10} and the nonlinear frequency responses~\cite{Barton10,Bureau13,Schilder15}, nonlinear normal modes~\cite{Renson16,RensonIMAC2016} and limit-point bifurcation curves~\cite{Renson17} of several simple systems subjected to external harmonic excitations. Following the same general principles as CBC, phase-locked loops were exploited to trace the nonlinear frequency response of a nonlinear oscillator in ~\cite{MojrzischPAMM}. A similar method was used to measure the backbone curves of a beam with nonlinear boundary conditions~\cite{Peter17} and of a circular plate and a Chinese gong~\cite{Denis18}.\\

The experimental demonstration of CBC remains largely limited to systems whose dynamics can be approximated by SDOF oscillators. The objective of this paper is to demonstrate experimentally that CBC is applicable to MDOF systems and useful to study their multi-mode dynamics. To this end, the mechanical structure investigated here is tuned to exhibit a 3:1 modal interaction. In the interaction region, strong nonlinear couplings exist between the first two modes of the system, such that it cannot be reduced to a SDOF oscillator. The presence of the modal interaction leads to an isola in the set of periodic response solutions at a given forcing amplitude. Numerical and experimental evidence of isola created by the presence of modal interactions were recently reported in~\cite{Kuether15b,Mangussi16} and~\cite{Shaw16,Gatti17,Detroux18}, respectively. Detecting and capturing the presence of such an isolated solution is important as missing them can lead to a significant underestimation of the resonance frequency and response amplitude of the system (as shown in Section~\ref{sec:rig}). \\ 

The structure of the paper is the following. In Section~\ref{sec:exp}, the experimental set-up considered in this paper is described (Section~\ref{sec:rig}) and the experimental results obtained using open-loop step sine tests are discussed (Section~\ref{sec:step-sine}). In Section~\ref{sec:intro_cbc}, the simplified CBC method used in the paper is introduced (Section~\ref{sec:cbc_sweeps}). This simplified approach dispenses with sophisticated continuation algorithms, which makes CBC straightforward to apply. The stabilising feedback controller is then briefly presented in Section~\ref{sec:id}. The controller is deliberately kept simple to show that CBC does not necessarily require sophisticated control strategies. In particular, linear control law is chosen and designed using just a linear model of the experimental set-up. The results obtained with CBC are presented in Section~\ref{sec:results}. It is shown that the isola can be systematically obtained by CBC without any a priori knowledge of its existence (Section~\ref{sec:results1}), which contrasts with the difficulties usually encountered with more-conventional test techniques such as stepped or swept sines. The invasiveness of the controller is analysed in Section~\ref{sec:discussion} and direct comparisons with open-loop experimental results obtained for similar excitation conditions are carried out in Section~\ref{sec:inva}. The effect of higher-harmonics in the force on the stability properties of the system is also investigated experimentally. The presence of such higher harmonics is well known in the literature~\cite{Claeys14,Chen16} but their effects on the dynamic behaviour of the system are rarely studied experimentally. Section~\ref{sec:conclusion} draws the conclusions of this study.

\section{The nonlinear experiment}\label{sec:exp}
\subsection{Description of the experimental set-up}\label{sec:rig}
\begin{figure}[t]
\centering
\begin{tabular*}{1.\textwidth}{@{\extracolsep{\fill}} c c}
\multicolumn{2}{c}{\subfloat[]{\begin{overpic}[scale=0.5,unit=1mm,width=0.8\textwidth]{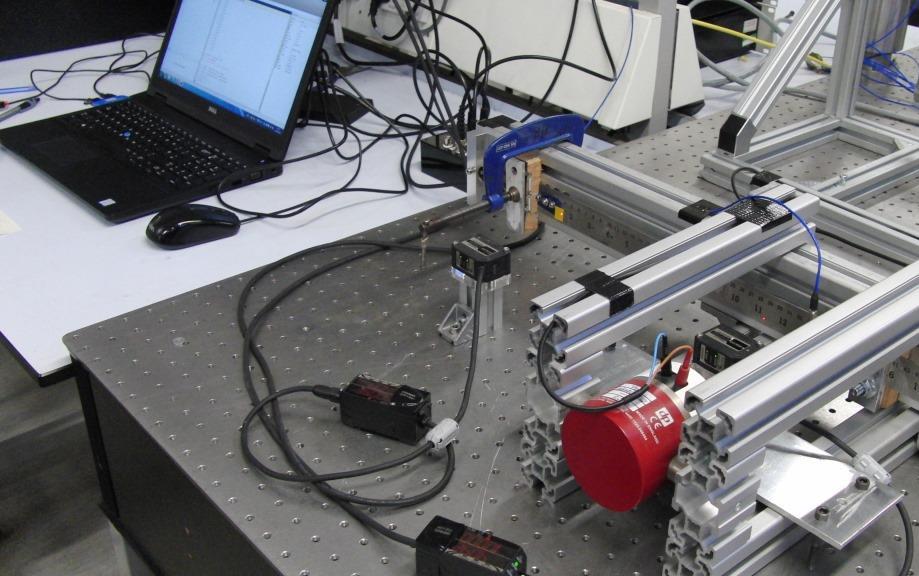}
%grid
\put(1,80){\colorbox{white}{Laptop}}
\put(50,80){\colorbox{white}{Real-time controller box}}
\put(77,68){\colorbox{white}{Nonlinearity}}
\put(97,60){\colorbox{white}{Beam}}
\put(77,2){\colorbox{white}{Shaker}}
\put(43.5,40){\colorbox{white}{Laser 1}}
\put(80,35){\colorbox{white}{Laser 2}}
\put(110,52){\colorbox{white}{Force sensor}}
\put(120,15){\colorbox{white}{Clamp}}
\linethickness{2pt}
\put(15,79){\color{white}\vector(1,-1.5){5}}
\put(70,79){\color{white}\vector(0,-1){11}}
\put(88,67){\color{white}\vector(-1,-1){8}}
\put(98,60){\color{white}\vector(-1,-1){8}}
\put(120,51){\color{white}\vector(0,-1){8}}
\put(85,5){\color{white}\vector(0.4,2){2}}
\put(58,41){\color{white}\vector(2,1){10}}
\put(125,18){\color{white}\vector(1,2.5){4}}
\put(95,36){\color{white}\vector(1,-0.1){10}}
\end{overpic}}}\\
\subfloat[]{\begin{overpic}[scale=0.5,unit=1mm,width=0.55\textwidth]{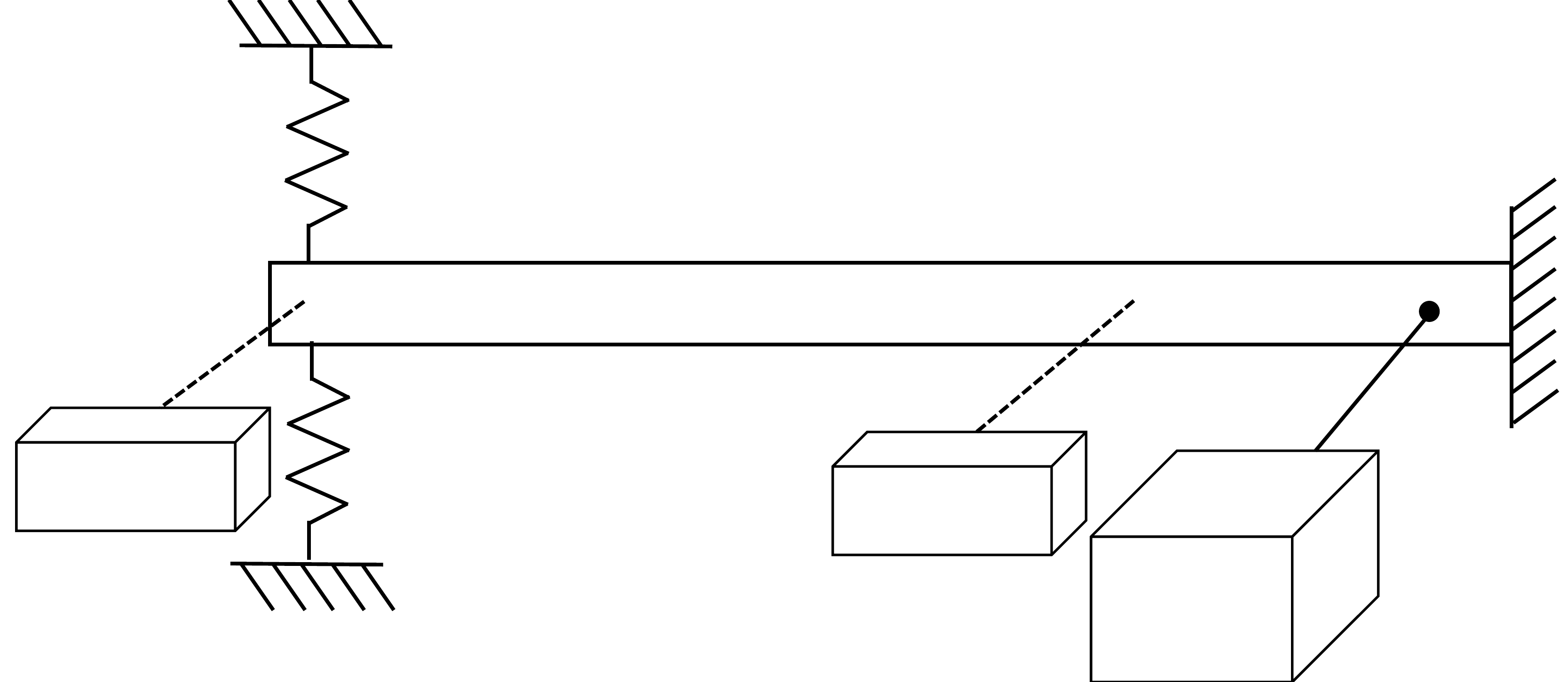}
\put(1,10){\small Laser 1}
\put(50,9){\small Laser 2}
\put(65.5,3.5){\small Shaker}
\put(60,30){\small Force sensor}
\put(80,29){\color{black}\vector(1,-1.3){4}}
\end{overpic}} & \subfloat[]{\label{nlfr1}\includegraphics[width=0.42\textwidth]{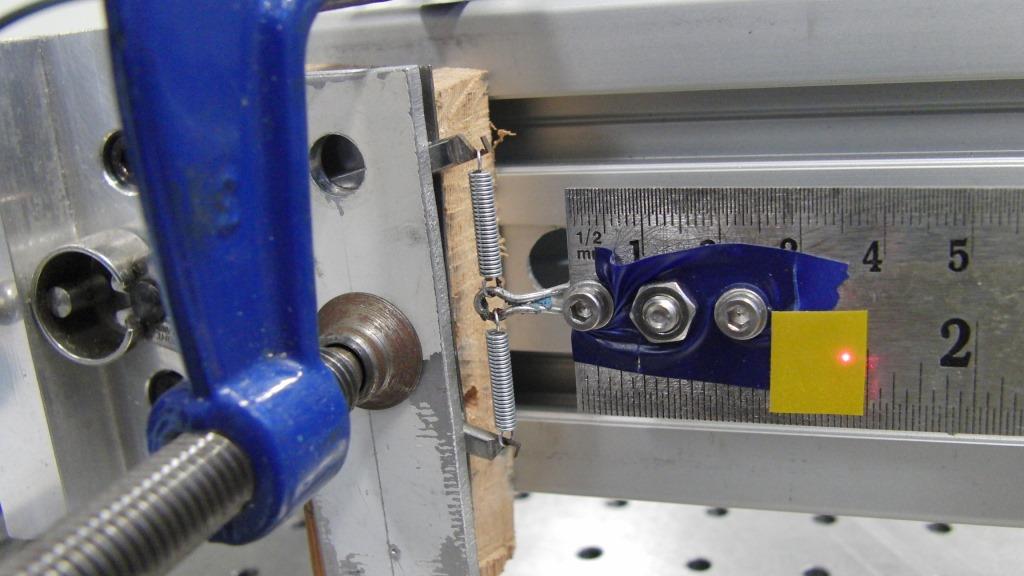}}
\end{tabular*}
\caption{Experimental set-up. (a) Overall experiment. (b) Schematic representation of the experiment. (c) Nonlinear mechanism at the free end of the cantilever beam.}
\label{fig:setup}
\end{figure}
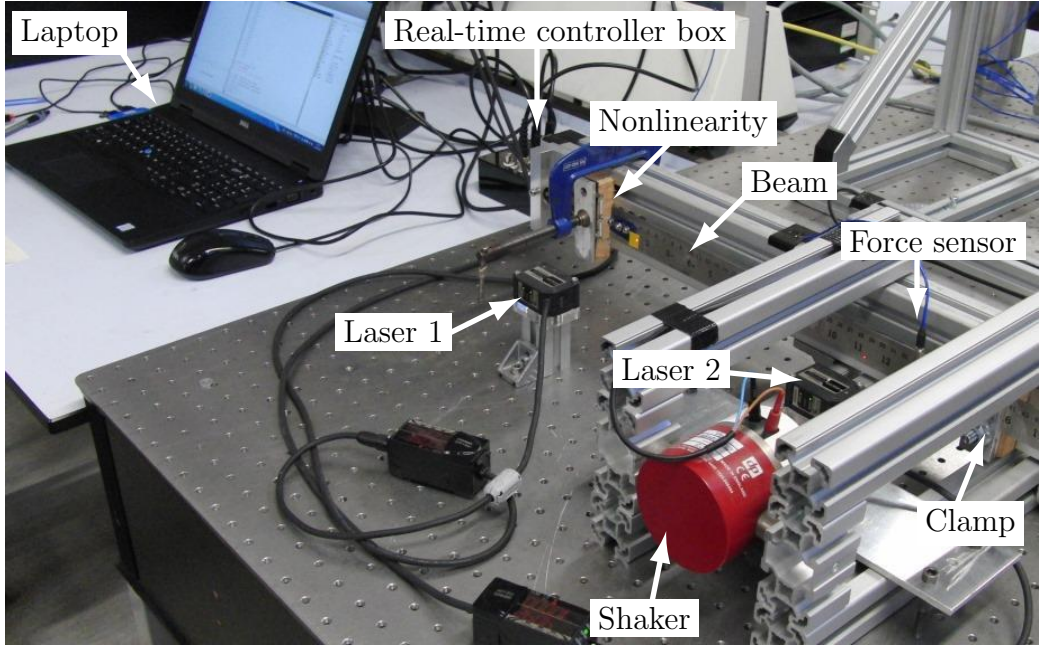
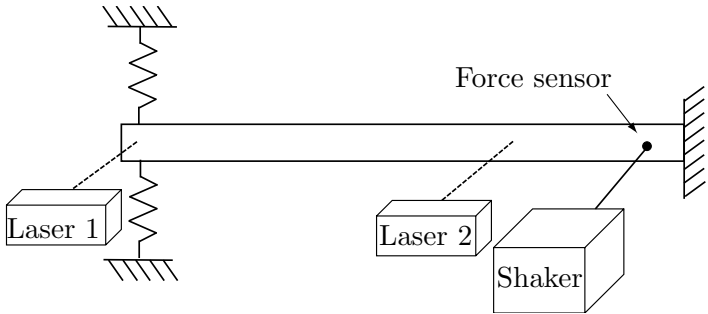
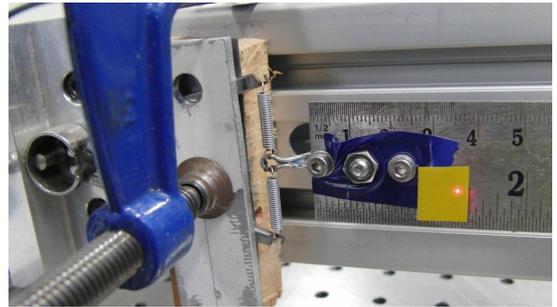

The experimental set-up considered here is shown in Figure~\ref{fig:setup}. The main structure is a cantilever beam of length 380 mm. The beam's free end is attached to two linear springs arranged as shown in Figure~\ref{fig:setup}(b). This mechanism gives rise to geometric nonlinearity at large amplitudes. Previous work has shown that the stiffness properties of this mechanism can be approximated by a linear plus cubic term~\cite{Shaw16}. However, a mathematical model of the nonlinearity is unnecessary for CBC. As such, neither the identification of the nonlinear parameters nor the exploitation of the mathematical form of the nonlinearity were necessary. The length of the beam as well as the pre-tension in the springs were carefully adjusted such that the ratio between the natural frequencies of the first two bending modes is close to, but larger than 3. This leads to the presence of a 3:1 modal interaction between these modes.\\

The structure is excited approximately 40 mm away from the clamp using a GW-V4 Data Physics shaker powered by a PA30E amplifier. The force applied to the structure is measured using a PCB208C03 force transducer connected to a Kistler signal conditioner (type 5134). The vibrations of the beam are measured at the tip and 55 mm away from the excitation using Omron ZX2-LD100 and ZX2-LD50  displacement laser sensors (lasers 1 and 2 in Figure~\ref{fig:setup}). The beam structure, the first laser sensor, the shaker and its power amplifier constitute the nonlinear experiment tested using CBC. Laser 2 is used only for display.\\

The algorithm used by the CBC method and presented in Section~\ref{sec:intro_cbc} is run on a laptop computer directly connected to the real-time controller (RTC) box via a USB cable. The RTC box consists of a BeagleBone Black on which the feedback controller used by CBC is implemented. Note that CBC algorithms do not run in real-time, only the feedback controller does (see Figure~\ref{fig:ctrl_loop}). The BeagleBone Black is fitted with a custom data acquisition board (hardware schematics and associated software are open source and freely available~\cite{CBC_hardware}). All measurements are made at 1\,kHz sampling with no filtering. Estimations of the Fourier coefficients of the response, input force, and control action are calculated in real time on the control board using recursive estimators~\cite{Renson16}; however, this was for convenience rather than a necessity. 

\subsection{Open-loop experiment: step-sine excitation}\label{sec:step-sine}
Open-loop tests were first performed using standard step-sine excitations to provide a basis for comparison with CBC. Data points were collected every 0.1 Hz in frequency after waiting three seconds for the transient dynamics to die out. The voltage signal to the shaker was a single-harmonic sinusoidal wave whose amplitude was kept constant. As such, the force applied to the structure varied with the excitation frequency, and was also found to include higher-harmonics.\\

\begin{figure}[htp]
\centering
\begin{tabular*}{1.\textwidth}{@{\extracolsep{\fill}} c c}
\subfloat[]{\label{step1}\includegraphics[width=0.45\textwidth]{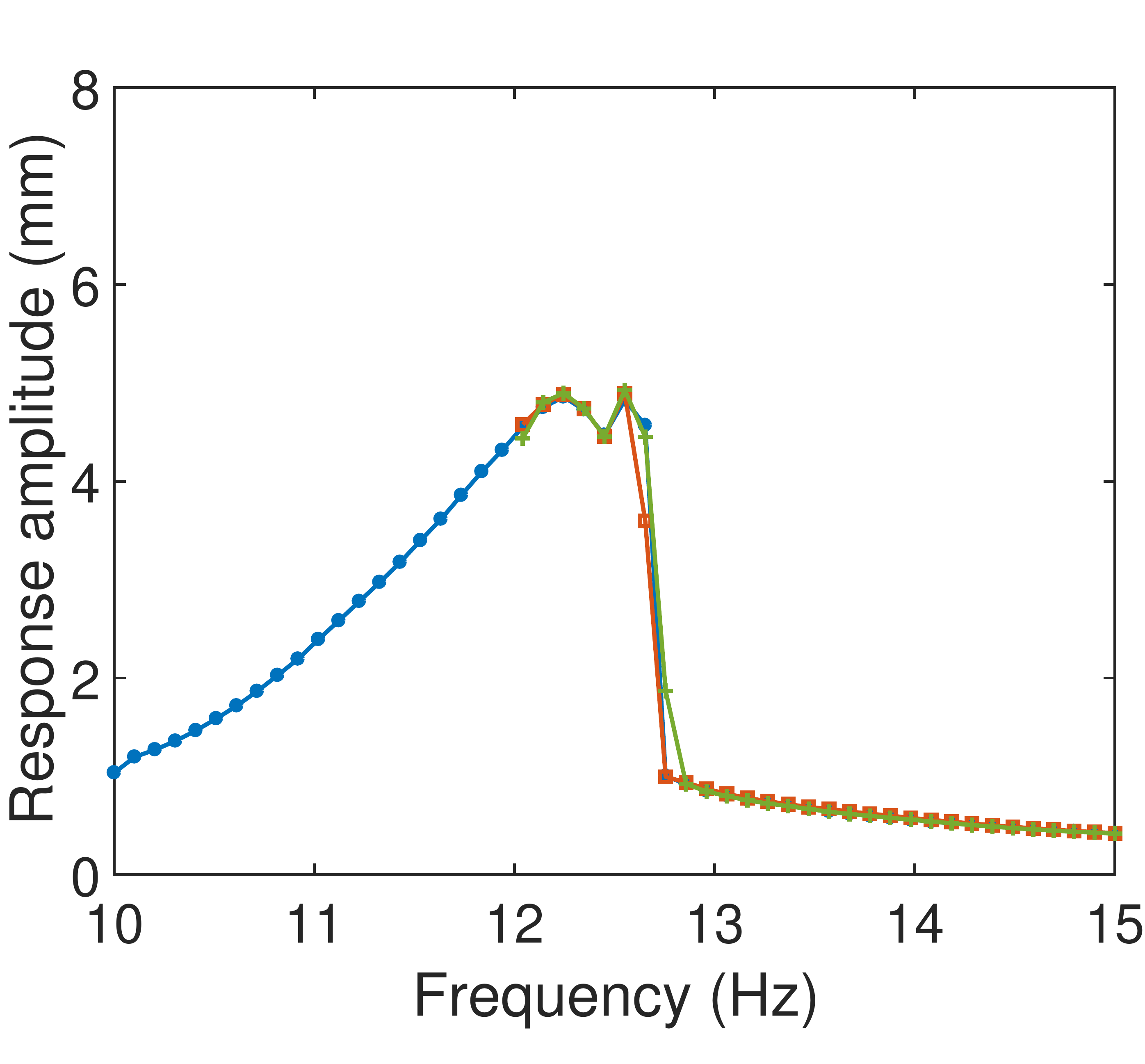}} &
\subfloat[]{\label{step2}\includegraphics[width=0.45\textwidth]{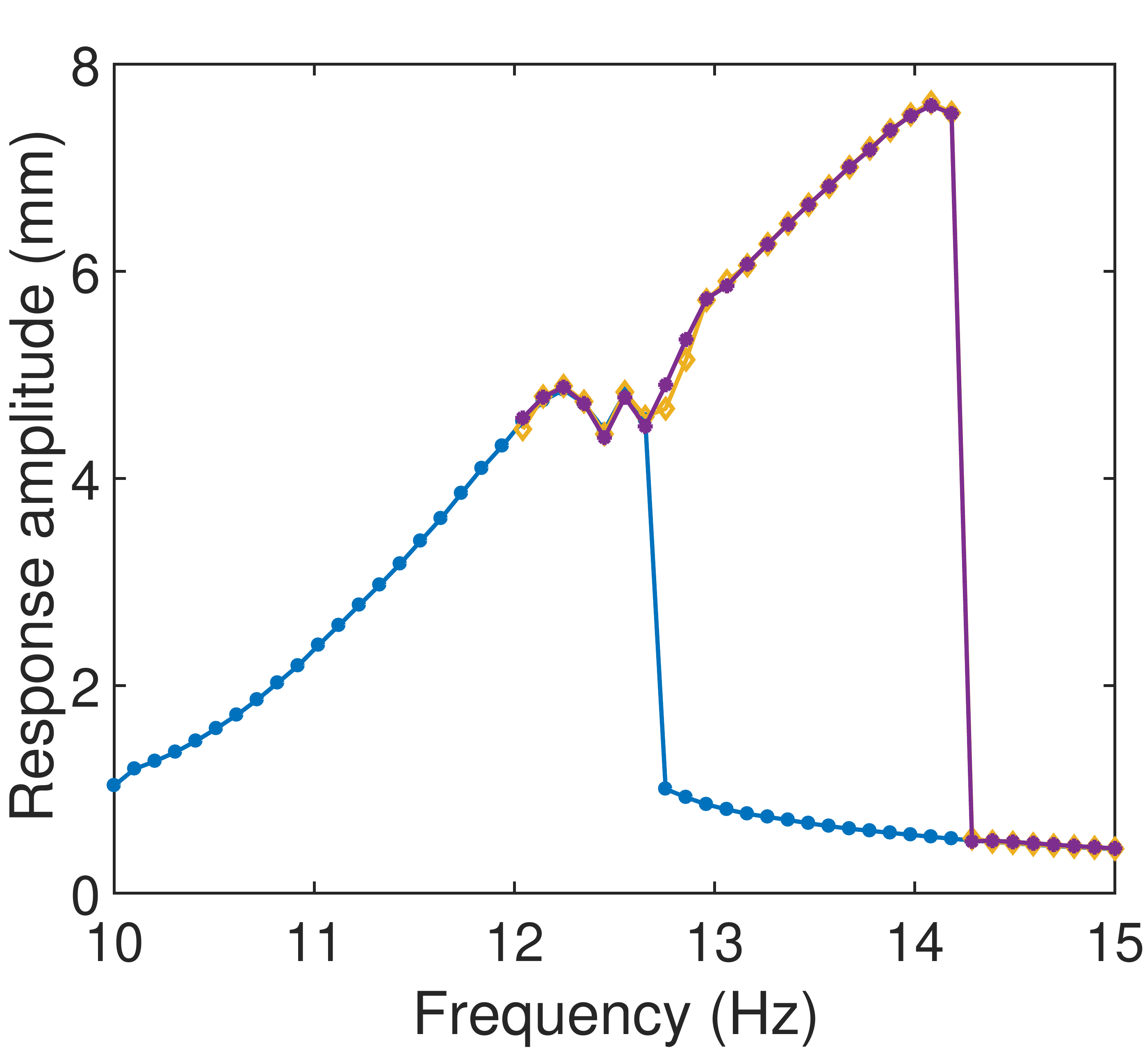}} \\
\subfloat[]{\label{qp1}\includegraphics[width=0.45\textwidth]{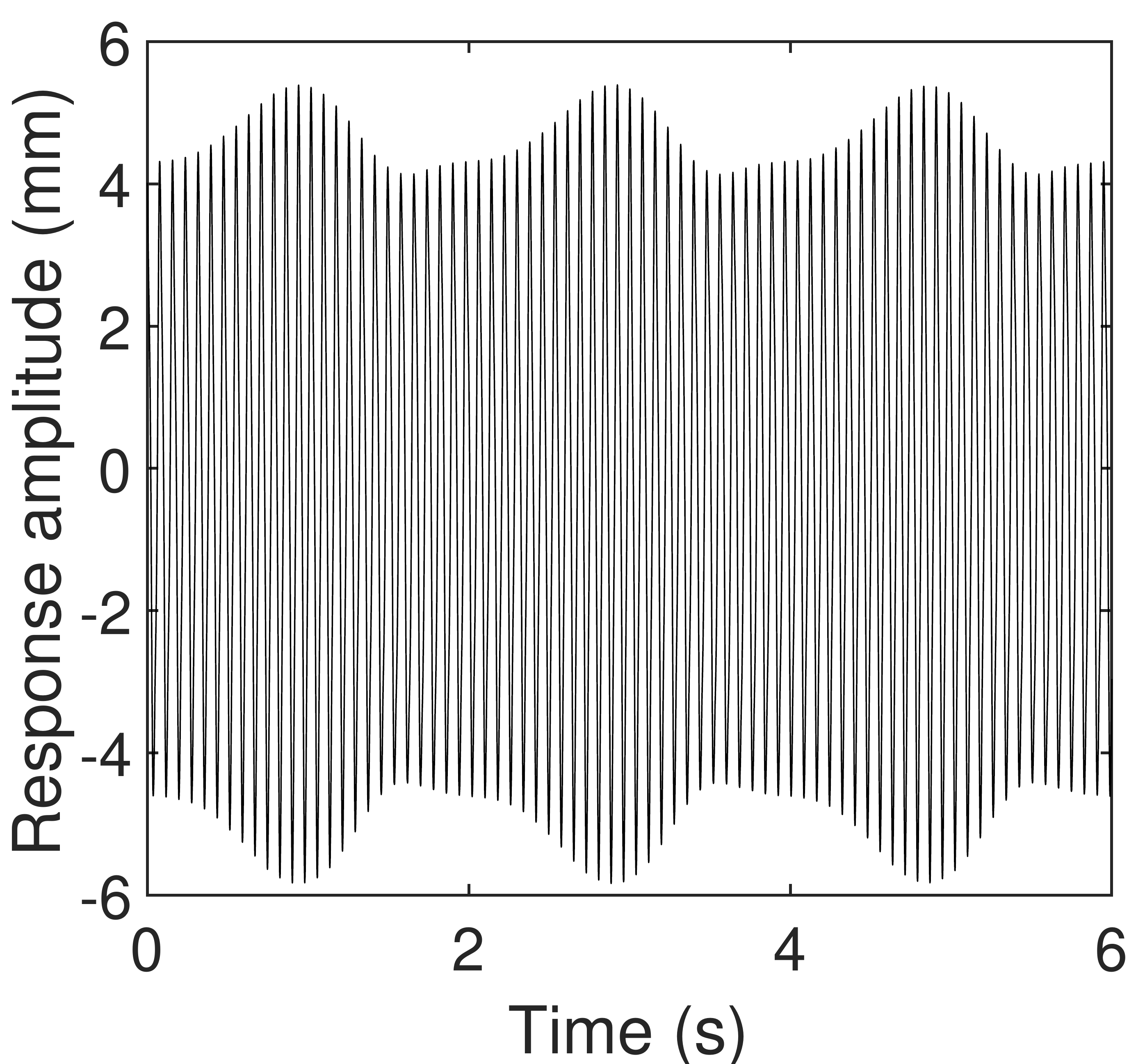}} &
\subfloat[]{\label{qp2}\includegraphics[width=0.45\textwidth]{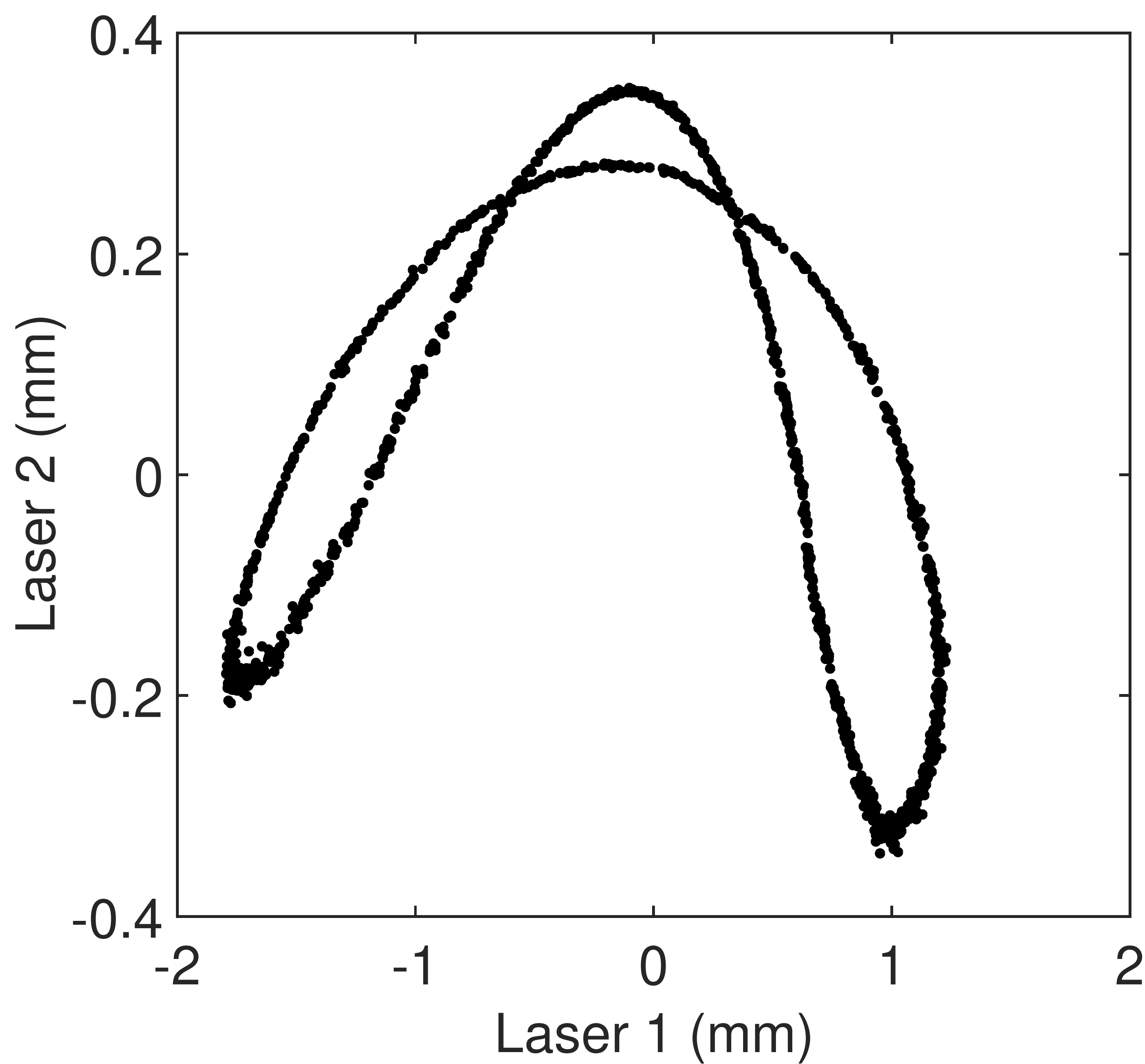}}
\end{tabular*}
\caption{Structure's response to step-sine excitation around its first natural frequency. (a, b) Markers (\textcolor{orange}{$\boldsymbol{\diamondsuit}$}, \textcolor{dgreen}{$\boldsymbol{+}$}, \textcolor{purple}{$\boldsymbol{*}$}, \textcolor{dred}{$\boldsymbol{\Box}$}) correspond to tests performed using identical parameters. The test performed between 10-15 Hz (\textcolor{dblue}{$\boldsymbol{\bullet}$}) serves to illustrate the overall shape of the resonance region. Branch of isolated periodic responses is captured only for the two tests  (\textcolor{orange}{$\boldsymbol{\diamondsuit}$}, \textcolor{purple}{$\boldsymbol{*}$}) presented in Subfigure (b). (c) Quasi-periodic oscillations observed at 12.6 Hz, i.e. between the resonance peak and the isolated solution. (d) Poincar\'e map of the time series presented in Subfigure (c).}
\label{fig:step_sine}
\end{figure}

Figures~\ref{fig:step_sine} shows the results of five consecutive tests performed with an identical excitation amplitude (0.3V). The first test (\textcolor{dblue}{$\boldsymbol{\bullet}$}), reported in both Figures~\ref{fig:step_sine}(a) and~(b), was performed between 10 Hz and 15 Hz and serves to illustrate the overall shape of the frequency response around the first natural frequency. The four other tests  (\textcolor{orange}{$\boldsymbol{\diamondsuit}$}, \textcolor{dgreen}{$\boldsymbol{+}$}, \textcolor{purple}{$\boldsymbol{*}$}, \textcolor{dred}{$\boldsymbol{\Box}$}) are identical and were performed between 12 Hz and 15 Hz. Figure~\ref{fig:step_sine}(b) shows that two tests (\textcolor{orange}{$\boldsymbol{\diamondsuit}$}, \textcolor{purple}{$\boldsymbol{*}$}) capture a branch of high-amplitude responses while the other two (\textcolor{dgreen}{$\boldsymbol{+}$}, \textcolor{dred}{$\boldsymbol{\Box}$}) drop at approximatively 12.5 Hz to capture lower-amplitude responses (Figure~\ref{fig:step_sine}(a)).\\

The high-amplitude stable periodic responses observed in Figure~\ref{fig:step_sine}(b) form a branch of responses that is disconnected from the main resonance peak. This \textit{isola} comes from the presence of a 3:1 modal interaction between the first and second bending mode of the structure as shown in~\cite{Shaw16} where the present structure was also investigated (without using CBC). Detecting and reaching an isola to capture the dynamics of the system there is a challenging task. A number of works have considered the application of perturbations, either in frequency or amplitude, from a coexisting, usually lower-amplitude, response~\cite{Shaw16,Detroux18}. For instance, in~\cite{Shaw16}, the isola was reached with a procedure that provided a brief period of very high amplitude forcing, before reducing force amplitude to the target level.\\

Here, the isola was reached due to the presence of a family of stable quasi-periodic oscillations bridging the gap between the main resonance peak and the isola. Figures~\ref{fig:step_sine}(c) and~(d) illustrate the quasi-periodic oscillations observed at 12.6~Hz. A clear modulation of the time series amplitude (measured at the beam tip) is observed in Figure~\ref{fig:step_sine}(c). A Poincar\'e map of the time series forms a closed curve, showing that the response is not periodic but quasi-periodic (Figure~\ref{fig:step_sine}(d)). For that particular level of excitation, the high-amplitude branch of stable periodic responses is considered to be isolated only with respect to other periodic responses. However, exploiting the presence of these quasi-periodic oscillations to reach the isola does not always work as observed in Figure~\ref{fig:step_sine}(a). Furthermore, this approach was also found to fail at other excitation levels. This is arguably due to the small size (or fractal structure) of the basin of attraction associated with the quasi-periodic oscillations and the possible presence of additional bifurcations.\\

For a rigorous testing process, the detection of important dynamic features such as an isola cannot be left to chance. Here, not capturing the isola leads to a significant underestimation of the resonance frequency and response amplitude by approximatively 14\% (1.9~Hz) and 35\% (2.7~mm), respectively. In simple systems the experiment can be randomly perturbed for each set of parameter values to try to reach different stable states, such as the isola. However, such a process is time consuming and does not scale to larger and more-complex structures. Section~\ref{sec:results} will show that CBC offers an effective and systematic alternative.

\FloatBarrier
\section{Introduction to control-based continuation}\label{sec:intro_cbc}
\subsection{Amplitude sweeps}\label{sec:cbc_sweeps}
In this paper, a simplified CBC method that dispenses with numerical continuation algorithms is exploited. The method is briefly explained here and the reader is referred to~\cite{Sieber10,Renson17} for additional details about the method.\\

\begin{figure}[t]
\centering
\begin{overpic}[scale=0.5,unit=1mm,height=6cm]{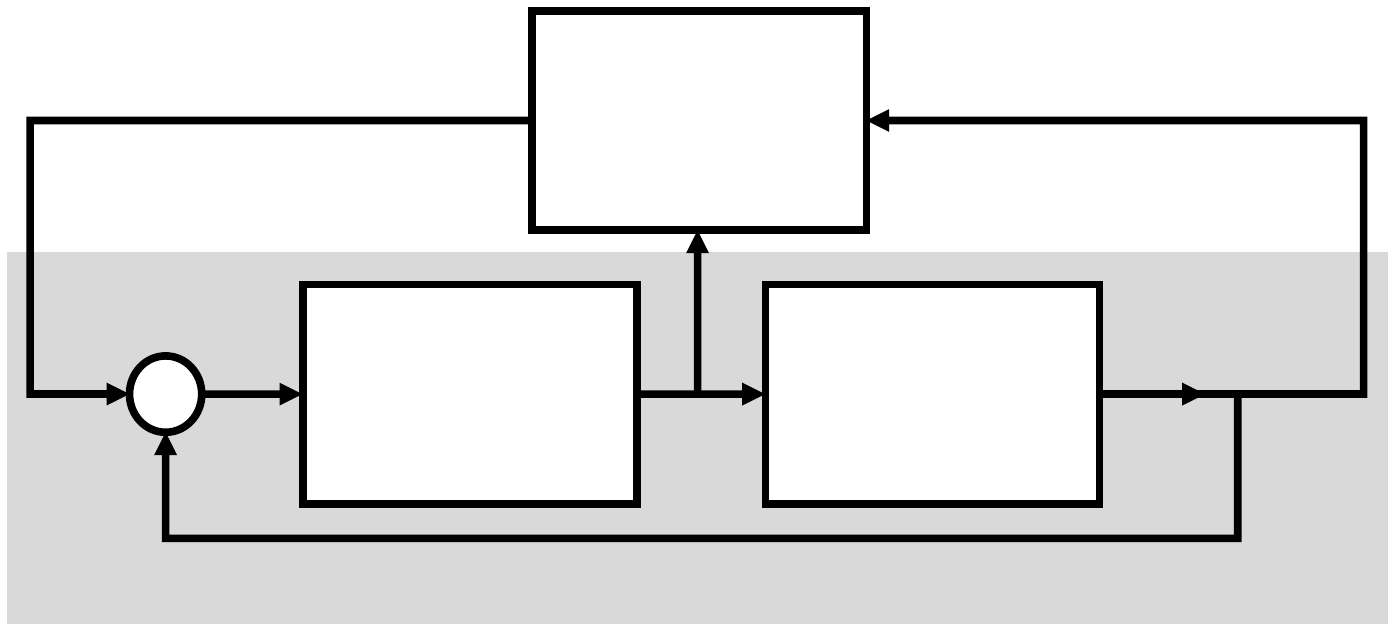}
%grid
\put(36,21.5){Controller}
\put(79,21.5){Experiment}
\put(62.15,51){CBC}
\put(57.75,45){algorithm}
\put(3.5,24.5){$x^*(t)$}
\put(10.8,16){$\boldsymbol{-}$}
\put(20,24){$e(t)$}
\put(63,18){$u(t)$}
\put(115,24){$x(t)$}
\put(57,2){\textit{\textbf{\textcolor{black}{Real-time}}}}
\end{overpic}
\caption{Schematic representation of a typical experiment performed using CBC. CBC algorithms do not operate in real-time and modify, on demand, the target signal $x^*(t)$ such that the controller becomes non-invasive.}
\label{fig:ctrl_loop}
\end{figure}

A schematic representation of a typical experiment performed using the simplified CBC method is shown in Figure~\ref{fig:ctrl_loop}. The experiment of interest is subject to a control signal $u(t)$ whose objective is to minimise the difference $e(t)$ between the response of the system $x(t)$ and a desired target signal $x^*(t)$. In the context of this experiment, $u(t)$ is the voltage signal to the shaker and $x(t)$ is the the tip displacement of the beam. No other external excitation is provided to the experiment.\\

Considering periodic motions, it is assumed that the response of the experiment and the control target can be decomposed into $m$ Fourier modes
\begin{equation}
x(t) = \frac{A_0}{2} + \sum^{m}_{j=1} A_j \cos (j \omega t) + B_j \sin (j \omega t) 
\quad \text{and} \quad
x^*(t) = \frac{A^*_0}{2} + \sum^{m}_{j=1} A^*_j \cos (j \omega t) + B^*_j \sin (j \omega t).
\end{equation}
Likewise, it is assumed that the control signal can be written in the same form 
\begin{equation}
u(t) = \underbrace{A^u_1 \cos (\omega t) + B^u_1 \sin (\omega t)}_{\text{fundamental}} + \underbrace{\frac{A_0^u}{2} + \sum^{m}_{j=2} A_j^u \cos (j \omega t) + B_j^u \sin (j \omega t)}_{\text{higher harmonics}}.
\label{eq:ctrl_sig_th}
\end{equation}

CBC seeks a reference signal $x^*(t)$ for which the control signal is non invasive, i.e. a control signal that does not modify the position in parameter space of periodic responses compared to the underlying uncontrolled experiment.\\

The simplified CBC method exploits the partition of the control signal in Eq.~\eqref{eq:ctrl_sig_th} to effectively reach such a non-invasive control signal at a constant forcing frequency. The fundamental harmonic component of the control signal is considered to apply a harmonic excitation to the experiment. The amplitude of this excitation can be viewed as a free parameter as the excitation is not fully determined by the user but depends on the response $x(t)$ and target $x^*(t)$. The higher-harmonics represent additional contributions that must be eliminated to recover the response of the experiment under pure harmonic excitation. The CBC algorithm minimises the amplitude of these higher-harmonics and can be viewed to act as a second, outer-loop control system acting on $x^*(t)$. For simplicity, this task is usually not performed in real-time but rather iteratively using Newton-like algorithms that modify $\bar{X}^* = (A^*_0,A^*_j,B^*_j)^m_{j=2}$ on demand.\\

\begin{figure}[ht]
\centering
\begin{overpic}[scale=0.5,unit=1mm,width=0.8\textwidth]{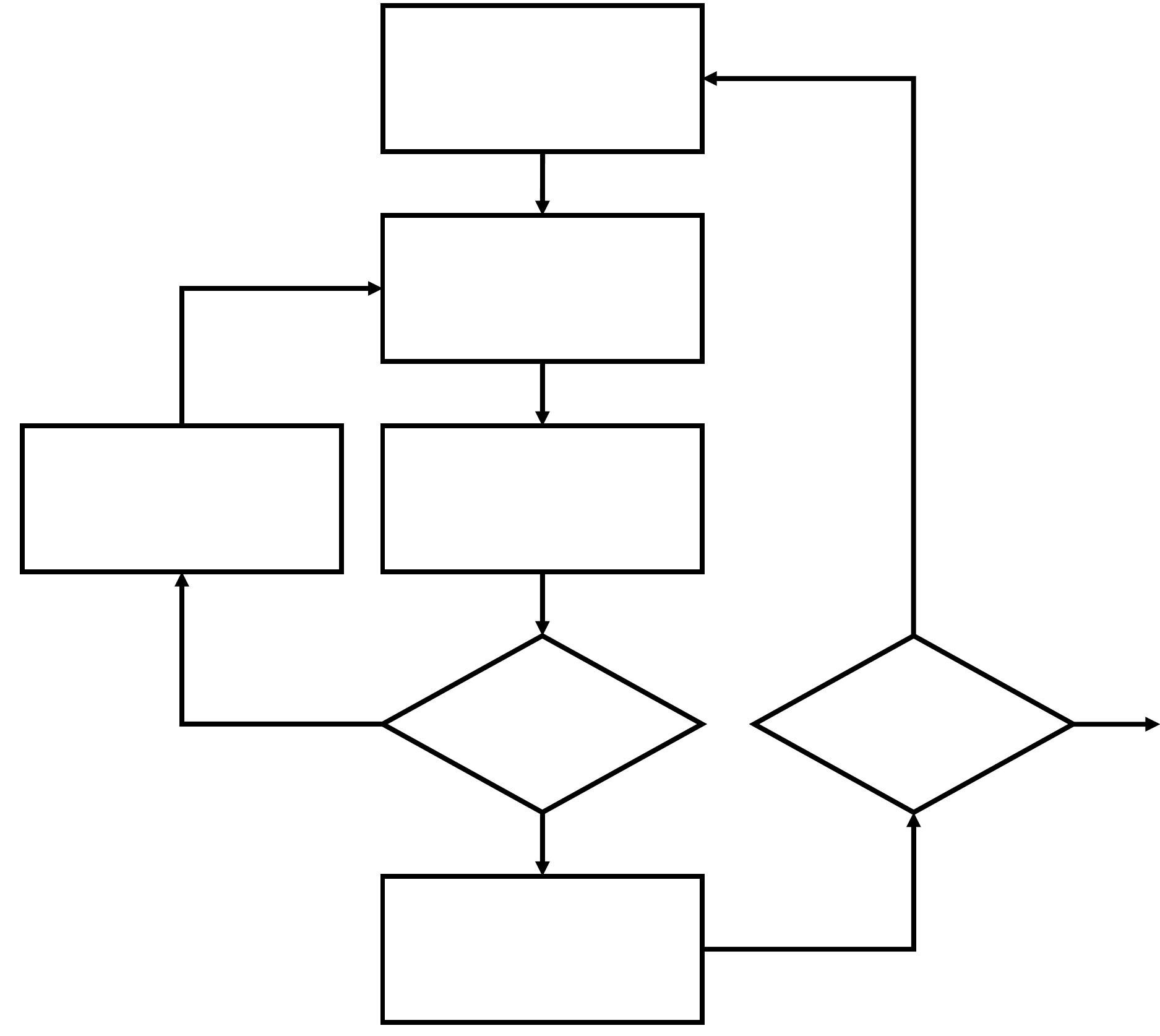}
%grid
\put(51,111){Increment $B^*_1$}
\put(46.25,89){Update controller's}
\put(55.5,84.5){demand}
\put(46.5,64){Wait for transients}
\put(54,59.5){to die out}
\put(7,66.5){Correct higher-}
\put(10.75,61.75){harmonics}
\put(6.5,57){$A^*_0,\;(A_j^*,B_j^*)^m_{j=2} $}
\put(53.5,35.5){$\left| \bar{U} \right|_{\infty} \leq \delta$}
\put(47,9){Record data point}
\put(32,37){FALSE}
\put(64,22.5){TRUE}
\put(96,36){$||X||_2 \geq x_{\text{max}}$}
\put(93,50){FALSE}
\put(125,37.5){END}
\end{overpic}
\caption{Block diagram describing the simplified CBC algorithm used in this paper. $A_1^* = 0$ throughout.}
\label{fig:cbc_algo}
\end{figure}

The algorithm of the method is schematicised in Figure~\ref{fig:cbc_algo}. Considering all the target coefficients to be initially equal to zero, it consists of the following steps:
\begin{enumerate}
\item Increment the fundamental target coefficient $B^*_1$ ($A^*_1$ is left equal to zero throughout).

\item Pass on the new target coefficients to the real-time controller, which updates the demand. 

\item Wait for the experiment to reach steady-state. The appropriate duration will not only depend on the difference between previous and new target coefficients but also on the characteristics of the controller. Note that, for this simplified algorithm, the difference between the response of the system and the target signal will never be zero as there is no external excitation other than the control signal.

\item Estimate the higher-harmonic coefficients $\bar{U}=\left(A^u_0, A^u_j, B^u_j\right)^{m}_{j=2}$ of the steady-state control signal. In the first instance, these coefficients are unlikely to be equal to zero due to harmonics in $x(t)$. In this case, the controller is invasive and the steady-state response reached is not a valid data point.

\item Correct the higher-harmonic target coefficients $\bar{X}^*=(A^*_0,A^*_j,B^*_j)^m_{j=2}$ in order to decrease $|\bar{U}|_{\infty}$ below a user-defined tolerance $\delta$. Here, this is effectively achieved using a Picard (fixed-point) iteration algorithm as in~\cite{Renson16}.
 
\item After convergence, record the obtained data point. The higher-harmonic coefficients of the control signal equal zero (up to experimental accuracy) and the fundamental coefficients $(A_1^u,B_1^u)$ represent the total, single-harmonic, input applied to the experiment. At this point, we can claim that the controller is noninvasive and that the position in parameter space of the observed periodic response is identical, within tolerance $\delta$, to the one in the underlying uncontrolled experiment with an excitation $u(t)=A_1^u \cos (j \omega t) + B_1^u \sin (j \omega t)$. 

\item Repeat the whole procedure (from step 1) until a desired maximum response of the system is reached.
\end{enumerate}

This algorithm can be viewed as an amplitude sweep that allows us to trace the evolution of the applied excitation (the free parameter) as a function of the response amplitude of the system. As shown in Section~\ref{sec:results}, the collected data points usually form S-shaped curves, which can be collected at different excitation frequencies and assembled to map the response of structure as a function of the excitation amplitude and frequency.\\

The number of Fourier modes $m$ should theoretically be infinite but, in reality, a limited number of harmonics is often sufficient to accurately describe the response of the system. Here $m$ is equal to 7 throughout the paper. In practice, the control signal $u(t)$ will also include non-harmonic components coming from, for instance, measurement noise. This aspect is further discussed in Section~\ref{sec:discussion}.\\

The above CBC is based on the input to the whole experiment which comprises the shaker amplifier, the shaker, the structure and the first laser (see Section~\ref{sec:rig}) rather than the force applied to the structure. This force can be measured using a force transducer and directly considered when analysing the results. Given the nonlinear nature of the tested structure, the force generated by the shaker generally includes higher harmonics. These higher-harmonics cannot be directly removed using the higher-harmonic content of the control signal, i.e. $\bar{U}$, without altering the dynamics of the underlying uncontrolled experiment or including their amplitudes as additional bifurcation parameters. However, the higher harmonics in the force can be compensate for by adding an external excitation signal that introduces higher harmonics at the input of the experiment~\cite{Renson17,Shaw16}. This harmonic-compensation procedure was not performed here.

\subsection{The stabilising feedback controller}\label{sec:id}
The challenge in CBC is to design a controller that can stabilise the open-loop experiment in the desired range of parameters. Although CBC is a model-free method, this task can be greatly facilitated by the knowledge of a mathematical model of the open-loop experiment.\\

In this study, a linear time invariant model of the open-loop experiment was identified using low-level broadband input-output data. The applied excitation consisted of a random multisine excitation, which is a periodic random excitation with a user defined amplitude-spectrum~\cite{Schoukens04}. Each sine component of the signal in the 5-70 Hz excitation band was given an identical non-zero amplitude and its phase was randomly drawn from a uniform distribution between $[0, \; 2\pi)$. The obtained time signal was scaled such that a signal with a root-mean-square value of 0.1V was applied to the shaker amplifier.\\

Collected experimental data were exploited to identify a single-input single-output state-space model of the experiment using a linear subspace identification technique. In particular, the \texttt{n4sid.m} function available in MATLAB's system identification toolbox was exploited. The five first periods of the signal were discarded due to the presence of transient dynamics and the last five periods were averaged prior to identification. Note that the input signal was considered to be the input voltage to the shaker, such that the identified model includes the effects of the dynamics of the excitation system.\\

Two-modes corresponding to the first two bending modes of the beam were clearly identifiable from the data. Their natural frequencies (damping ratios) were estimated at 11.49 Hz (0.026) and 36.45 Hz (0.022), respectively. The identified model showed a 97\% fit to the data according to the normalised root mean square error criterion. Validation of this model on experimental data not used in the identification and collected with other realisations of the random phases confirmed this level of accuracy. The main source of error on the model comes from the nonlinear distortions present even for low levels of excitation.\\

The identified linear model proved sufficient to design the stabilising feedback controller necessary to CBC. In particular, pole placement techniques were used to introduce additional poles and modify the asymptotes of the root-locus diagram, allowing to consider larger gains. The locations of the additional poles was chosen such that the root-locus diagram appeared relatively insensitive to small errors in the identified model of the experiment. The discrete-time transfer function of the \textit{Controller} block (Figure~\ref{fig:ctrl_loop}) was chosen to be:
\begin{equation}
C(z) = \frac{0.0053}{z^3-2.4521 z^2+1.9725 z - 0.5155}.
\label{eq:controller}
\end{equation}
Other controllers possessing better time and/or frequency characteristics could have been considered. However this was not necessary for CBC to work as the method requires only a stabilising controller. This was found to be the case for the controller described in Eq.~\eqref{eq:controller} throughout the range of parameters considered in this study. 

\section{Experimental results obtained using CBC}\label{sec:results}
The conventional open-loop experiments described in Section~\ref{sec:step-sine} revealed the existence of an isola purely by chance. In this section, we show how CBC (as described in Section~\ref{sec:intro_cbc}) can be applied systematically to fully explore the dynamics of the MDOF system shown in Figure~\ref{fig:setup}. Results are presented in Section~\ref{sec:results1} and the invasiveness of the control signal is discussed in Section~\ref{sec:discussion}. Open- and closed-loop results are compared in Section~\ref{sec:inva}.

\subsection{Amplitude and frequency response curves}\label{sec:results1}
\begin{figure}[th]
\centering
\includegraphics[width=0.45\textwidth]{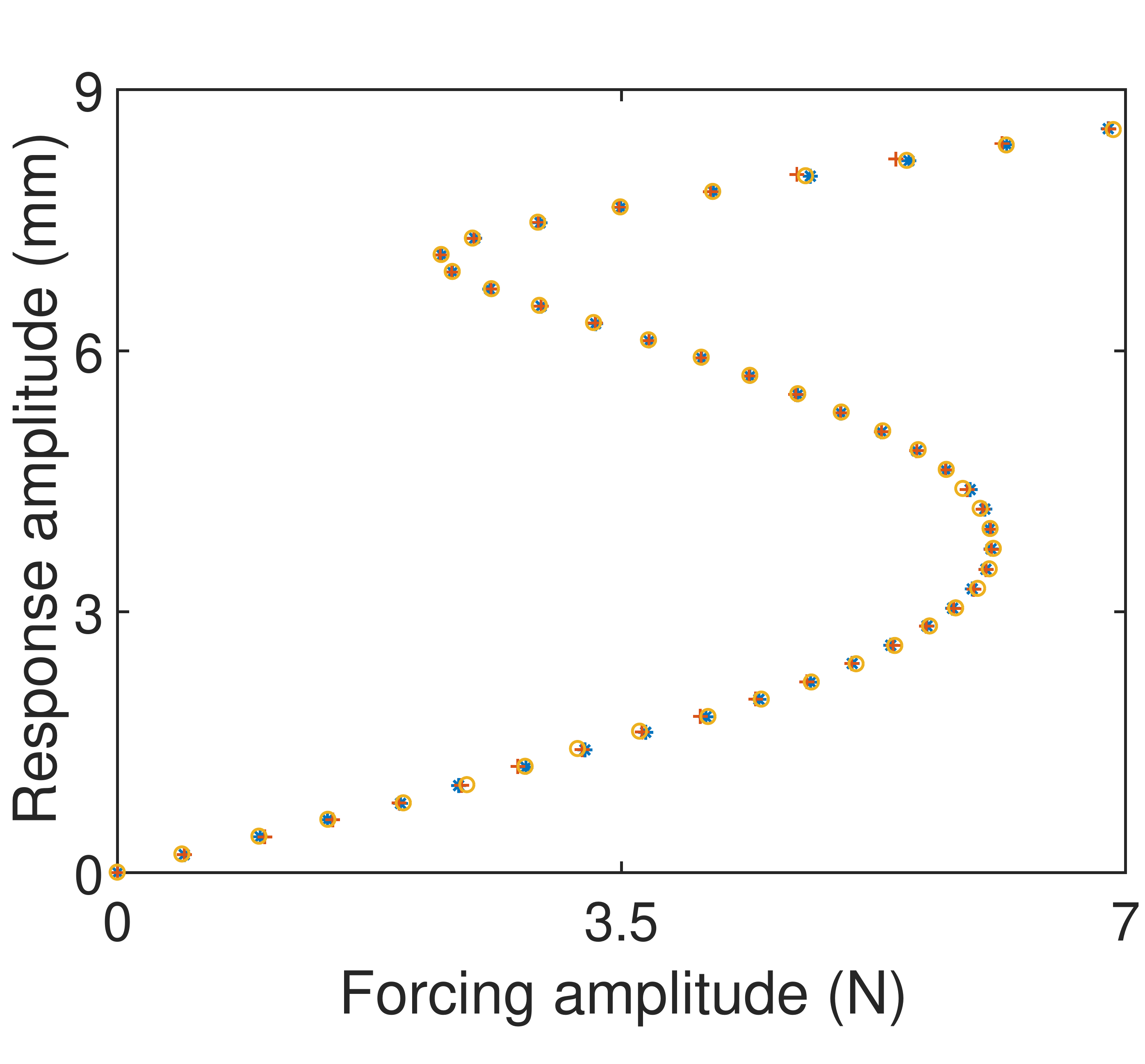}
\caption{Response curves measured at 13.5 Hz exhibit a S-like shape. Markers (\textcolor{dyellow}{$\boldsymbol{+}$}, \textcolor{dorange}{$\boldsymbol{+}$}, \textcolor{dblue}{$\boldsymbol{*}$}) correspond to different runs of the same test and serve to demonstrate the repeatability of the experimental results.}
\label{fig:Scurve_repeat}
\end{figure}

The evolution of the response amplitude of the beam tip as a function the amplitude of the applied force forms a characteristic S-curve, as shown in Figure~\ref{fig:Scurve_repeat}. An excitation frequency of 13.5 Hz and 0.2mm amplitude increments to the fundamental target coefficient were considered to generate this figure. Overall, the repeatability of the results is excellent as shown in Figure~\ref{fig:Scurve_repeat} where the markers (\textcolor{dyellow}{$\boldsymbol{+}$}, \textcolor{dorange}{$\boldsymbol{+}$}, \textcolor{dblue}{$\boldsymbol{*}$}) correspond to different runs of the same test. Some data points present a larger variability in the amplitude of the applied force, and this is due to slightly different higher harmonic content in the force.\\

Note that amplitudes presented in Figure~\ref{fig:Scurve_repeat} and thereafter include the contribution of the first seven Fourier modes. Those coefficients were estimated after the tests using time series containing 2000 samples, or between 22 and 29 oscillation cycles.\\

\begin{figure}[htbp]
\centering
\includegraphics[width=0.9\textwidth]{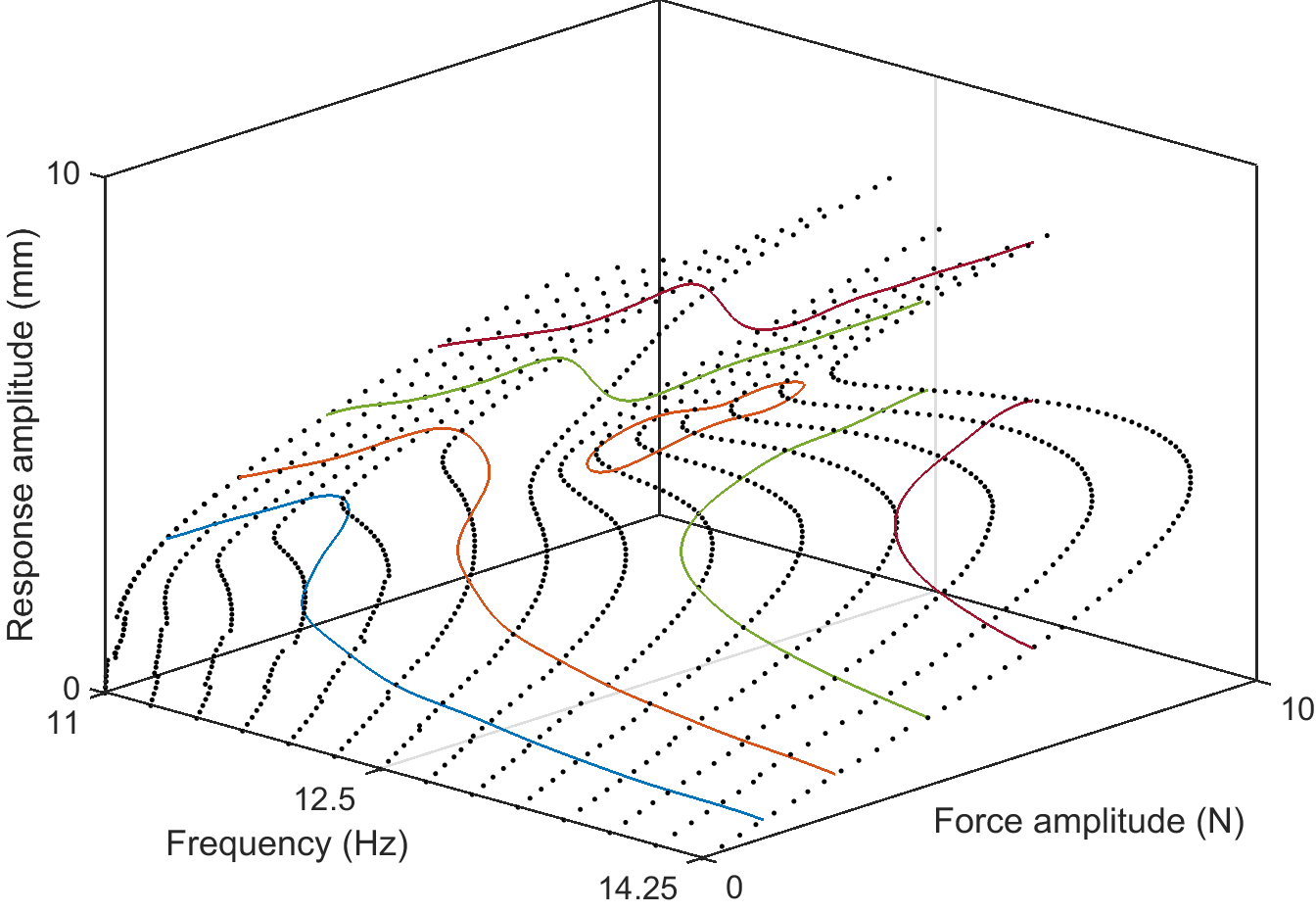}
\caption{Response of the beam structure as a function of the harmonic excitation amplitude and frequency. (\textcolor{black}{$\boldsymbol{\bullet}$}) Experimentally-measured data point. (\textcolor{dblue}{$\boldsymbol{-}$}, \textcolor{dorange}{$\boldsymbol{-}$}, \textcolor{dgreen}{$\boldsymbol{-}$}, \textcolor{dred}{$\boldsymbol{-}$}) Curve of constant force amplitude obtained using regression and numerical continuation.}
\label{fig:3Dresp}
\end{figure}

The response of the structure as a function of the excitation amplitude and frequency is mapped out as a sequence of S-curves. There are collected every 0.25 Hz between 11 Hz and 14.25 Hz and shown in Figure~\ref{fig:3Dresp} (\textcolor{black}{$\boldsymbol{\bullet}$}) where four numerically-generated nonlinear frequency responses curves (\textcolor{dblue}{$\boldsymbol{-}$}, \textcolor{dorange}{$\boldsymbol{-}$}, \textcolor{dgreen}{$\boldsymbol{-}$}, \textcolor{dred}{$\boldsymbol{-}$}) have also been added to help with the visualisation of the response surface.\\

The S-curve obtained at 12.5 Hz differs from the other curves by including an additional inflection point in its upper part. In fact, this curve marks a change in the form of the response surface of the system. A drop in the response amplitude can be observed in the upper part of the S-curves obtained right after 12.5 Hz. The forcing amplitude at which the S-curves bend back towards larger forcing amplitudes also starts to decrease after 12.5 Hz, which leads to the presence of the isola observed in Figure~\ref{fig:step_sine}(b).\\

Visual inspection of the response surface also reveals that some S-curves present a small but clear region where data points are shifted in force amplitude compared to the rest of the curve. This effect is particularly noticeable between 11 Hz and 12.25 Hz and seem to appear for a specific range of response amplitudes. The reasons behind this repeatable feature are unknown, but are not important for this study.\\

The nonlinear frequency response curves shown in Figure~\ref{fig:3Dresp} were obtained using a similar approach to the one used in~\cite{Renson16,Renson17}. In particular, a model of the response surface was built using Gaussian process (GP) regression and exploiting the unique parametrisation of the amplitude of the fundamental force component in terms of the response amplitude and excitation frequency. Curves of constant force amplitude that is, frequency response curves (\textcolor{dblue}{$\boldsymbol{-}$}, \textcolor{dorange}{$\boldsymbol{-}$}, \textcolor{dgreen}{$\boldsymbol{-}$}, \textcolor{dred}{$\boldsymbol{-}$}), were then calculated with numerical continuation applied to the GP model.\\
\begin{figure}[tbp]
\centering
\begin{tabular*}{0.9\textwidth}{@{\extracolsep{\fill}} c c}
\subfloat[]{\label{nlfr1}\includegraphics[width=0.45\textwidth]{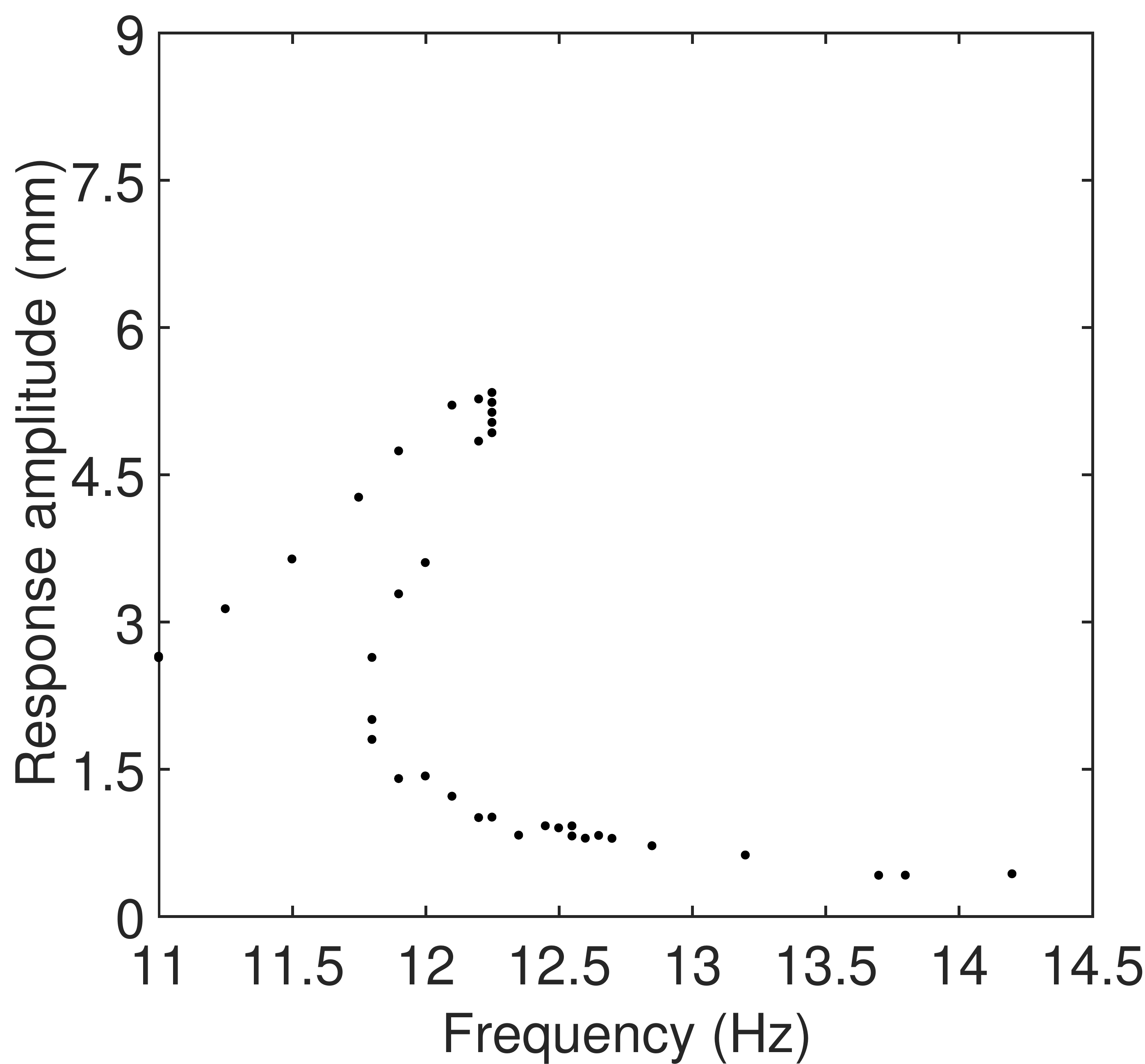}} &
\subfloat[]{\label{nlfr2}\includegraphics[width=0.45\textwidth]{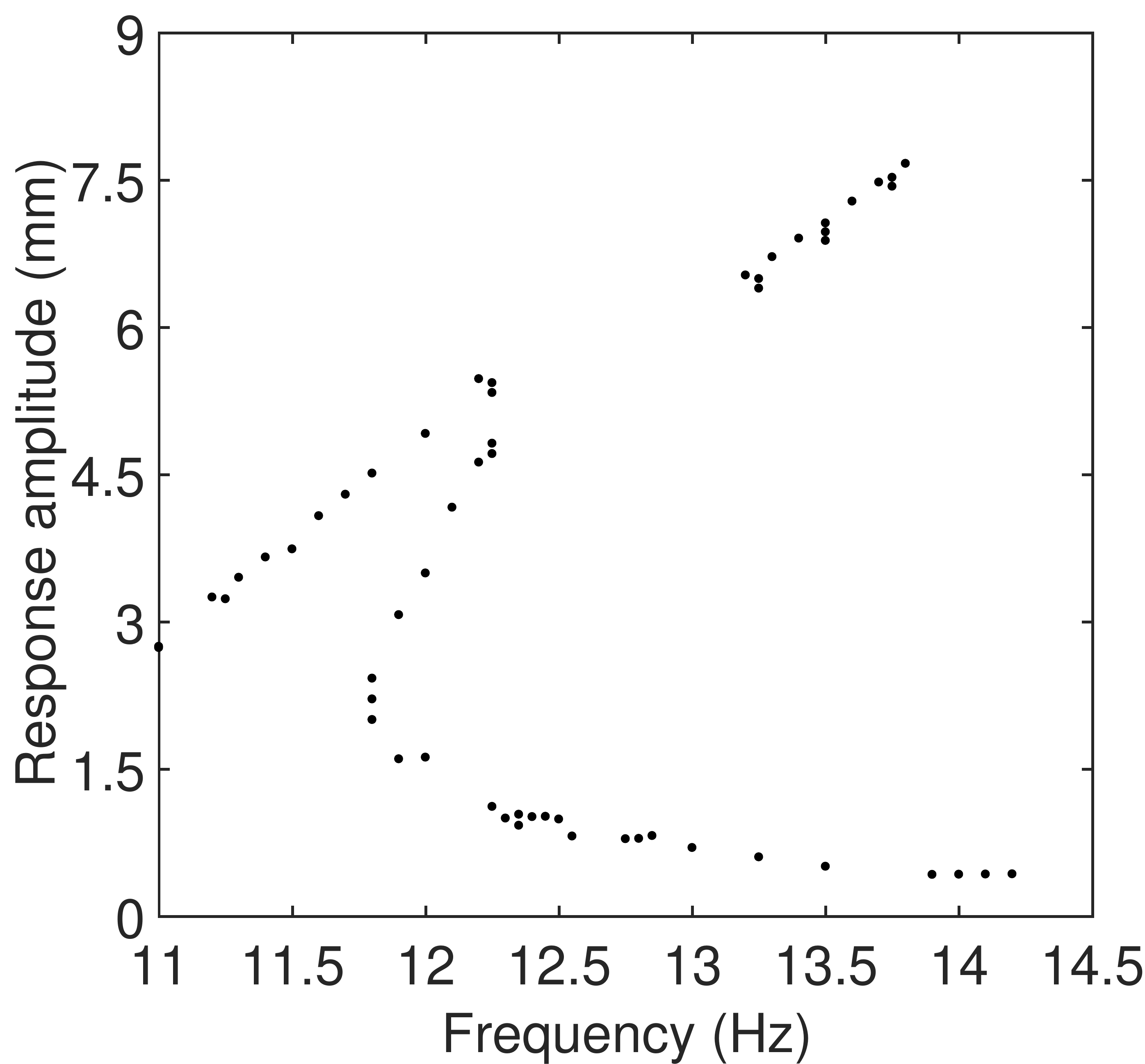}} \\
\subfloat[]{\label{nlfr3}\includegraphics[width=0.45\textwidth]{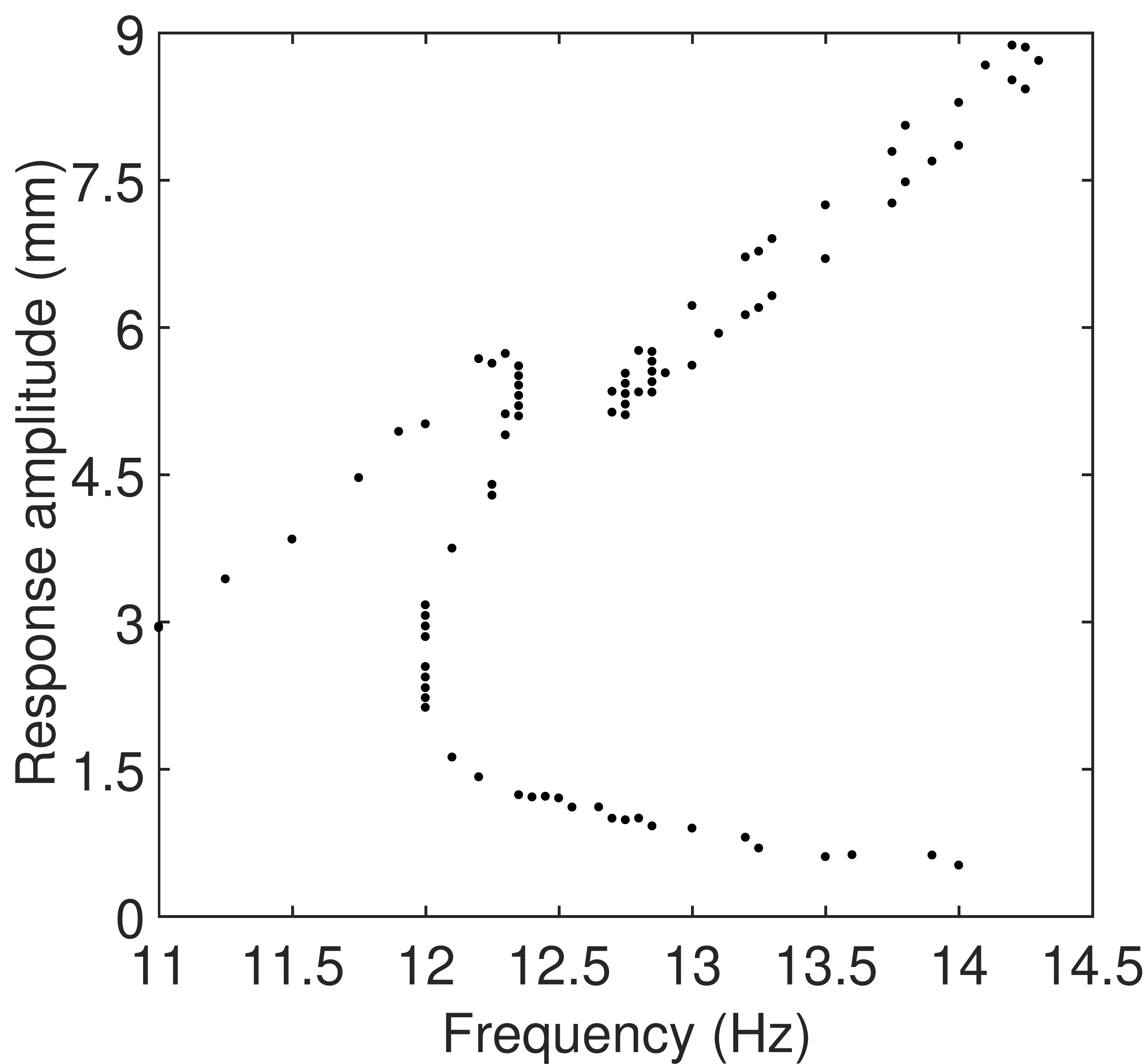}}&
\subfloat[]{\label{nlfr4}\includegraphics[width=0.45\textwidth]{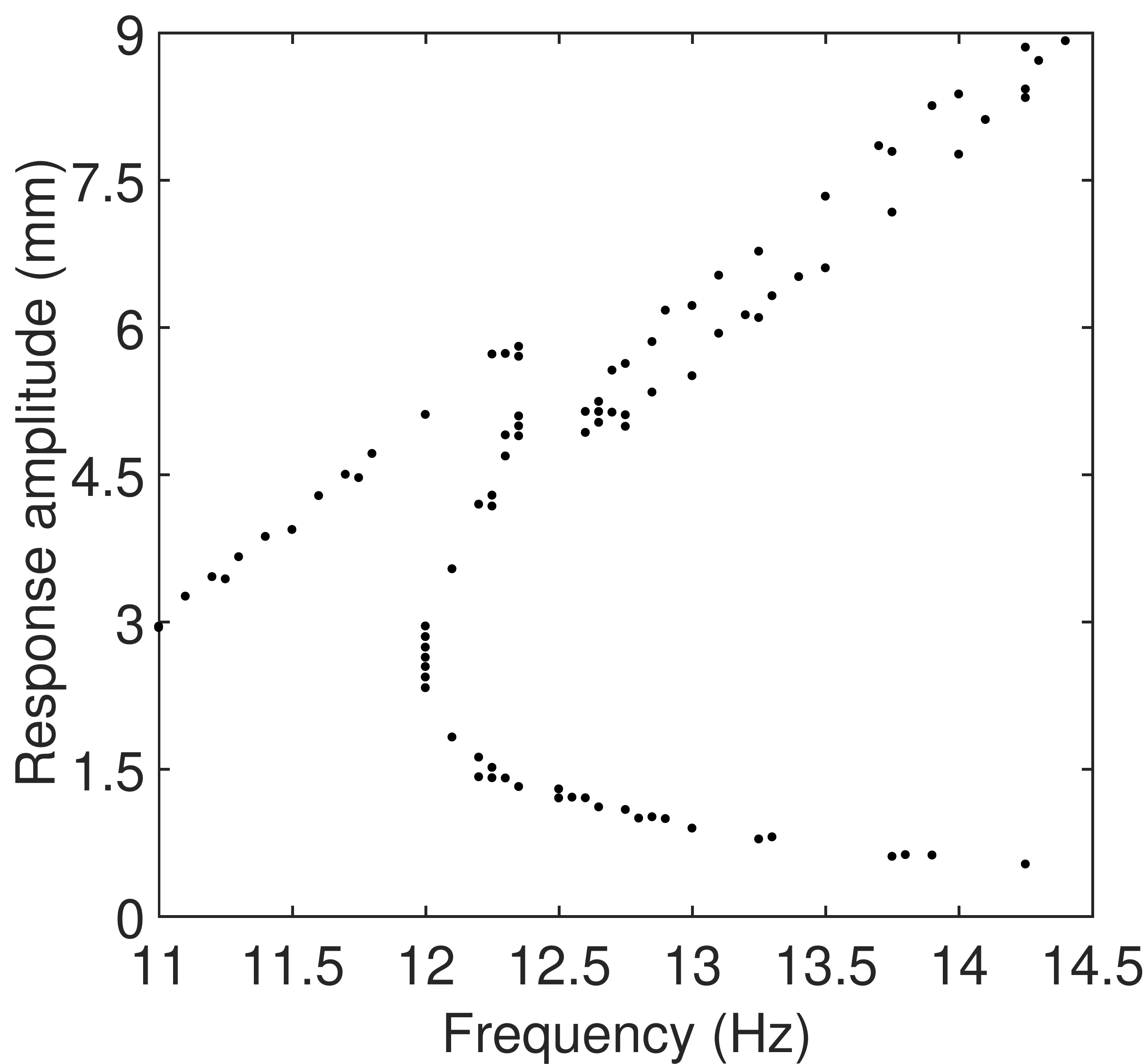}} \\
\subfloat[]{\label{nlfr5}\includegraphics[width=0.45\textwidth]{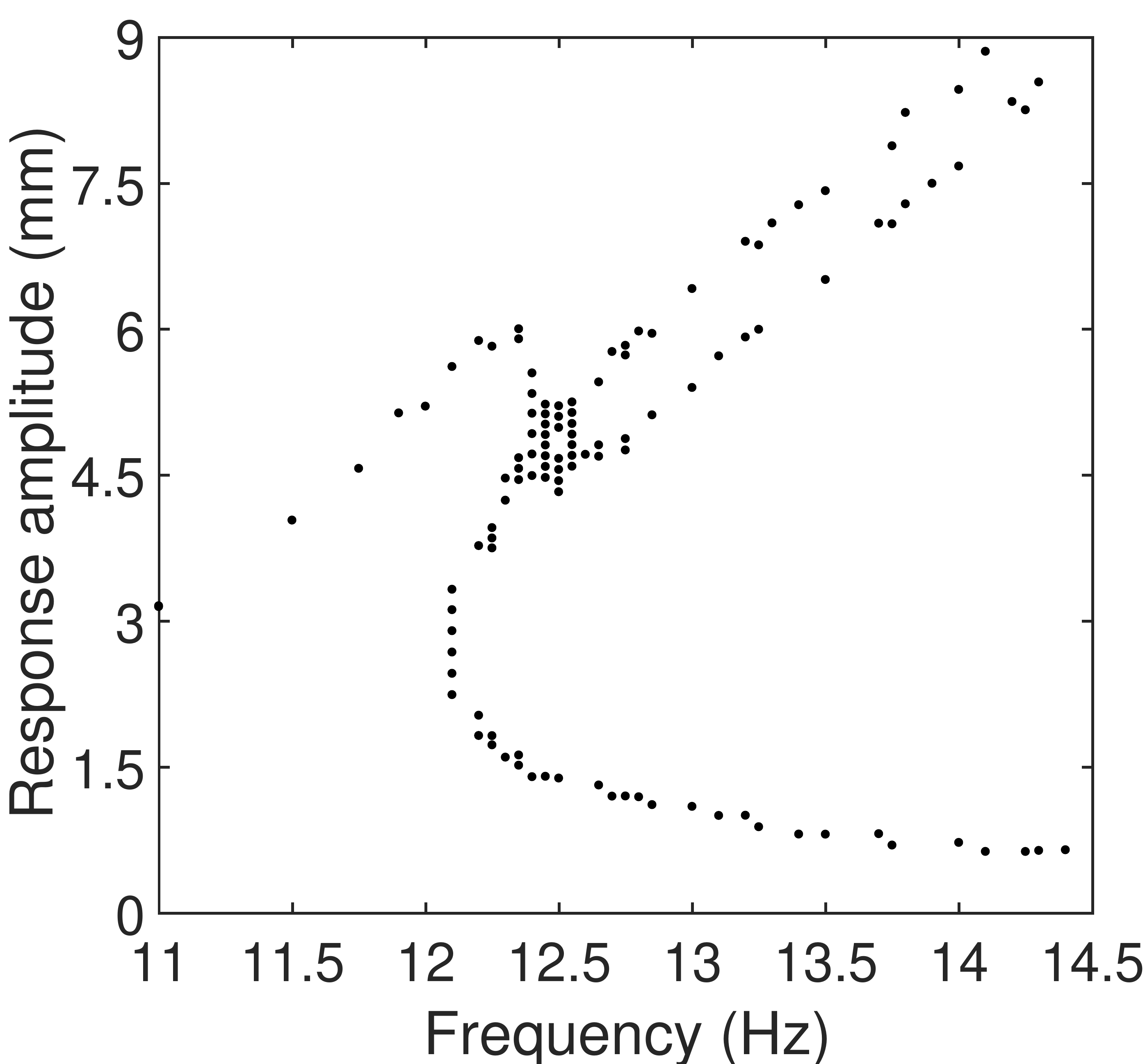}} &
\subfloat[]{\label{nlfr6}\includegraphics[width=0.45\textwidth]{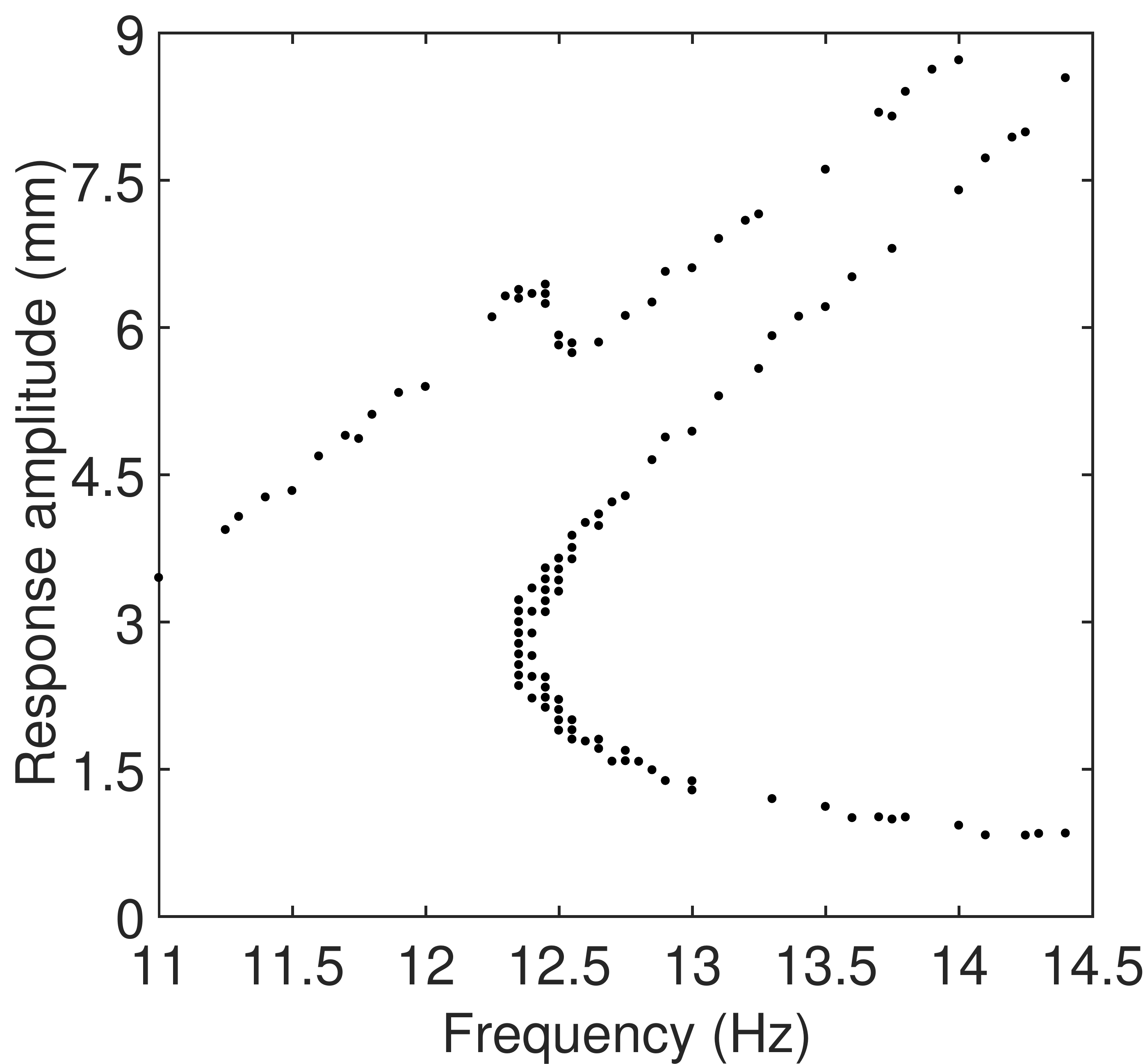}}
\end{tabular*}
\caption{Nonlinear frequency response of the beam structure obtained for different amplitudes of the harmonic excitation. (a-f) 1.1, 1.2, 1.5, 1.6, 1.85, 2.5 N, respectively.}
\label{fig:nlfr_slices}
\end{figure}

Regression techniques can create artefact features such as the undulations of the isolated curve observed in Figure~\ref{fig:3Dresp} (also see~\cite{Renson17}). To visualise the nonlinear frequency response of the system directly on the raw data, a simpler, qualitative, approach selecting the data points that correspond approximatively to a given force amplitude can be considered. Six nonlinear frequency response curves obtained in this way for 1.1N, 1.2N, 1.5N, 1.6N, 1.85N and 2.5N are shown in Figure~\ref{fig:nlfr_slices} (a-f), respectively. Data points found within $\pm$3\% of a selected force level were considered to be part of the same curve. Additional S-curves were collected to improve the frequency resolution of the figures. At 1.2N (Figure~\ref{fig:nlfr_slices}(b)), an isolated branch of high-amplitude responses disconnected from the main resonance peak appears. This isola grows as the forcing amplitude increases and eventually merges with the main resonance peak, leading to an abrupt change in the resonance frequency and response amplitude of the main branch. This step change was numerically observed on a beam system in~\cite{Kuether15}.\\

\begin{figure}[tbp]
\centering
\includegraphics[width=0.45\textwidth]{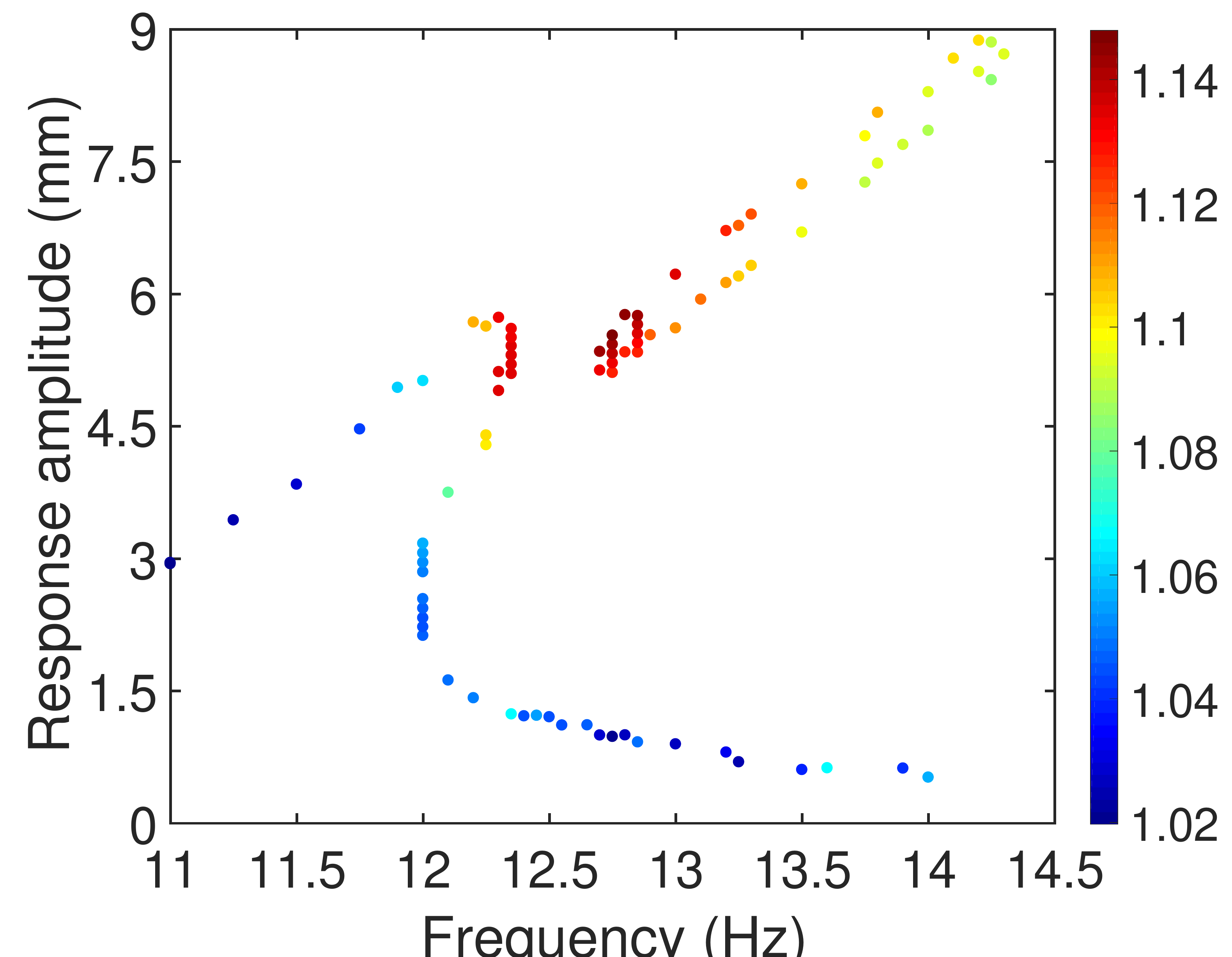}
\caption{Nonlinear frequency response of the system obtained at 1.5N. Data points are coloured according to the ratio between the amplitude of the third and fundamental harmonic components of the response.}
\label{fig:h3_nlfr}
\end{figure}

The existence of the isola is associated with the presence of a modal interaction between the first and second bending modes of the system~\cite{Shaw16}. This can be visualised in Figure~\ref{fig:h3_nlfr} where data points have been coloured according to the ratio between the amplitude of the third and fundamental harmonic components of the response. The region where the third harmonic is the largest is limited to the neighbourhood of 12.5 Hz; the high-amplitude region where the gap between the main resonance peak and the isola is observed and where the interaction is expected~\cite{Shaw16}. Away from this region, the frequency ratio between the two modes becomes increasingly larger than 3 and the interaction progressively decreases. As such, the magnitude of the third harmonic also decreases and becomes comparable to the amplitude of the other harmonics (not shown for conciseness).\\

This section has shown that CBC is a systematic method that can be used to characterise the dynamics of a nonlinear system. Compared to classical open-loop testing techniques such as stepped or swept sines, CBC provides a more complete picture of the system's dynamics, which facilitates the interpretation of the results. In particular, CBC was able to capture the isola present in the frequency response of the system. No a priori knowledge about its existence nor about the nonlinear dynamics of the system was necessary.

\newpage
\subsection{Invasiveness of the control signal}\label{sec:discussion}
The results presented in Section~\ref{sec:results1} were obtained under closed-loop experimental conditions. To guarantee that no artificial features were created by the CBC method, the results should be validated. This section analyses the invasiveness of the control signal by comparing its magnitude to the excitation applied to the experiment. Direct comparisons between closed- and open-loop experimental results obtained for similar excitation conditions are carried out in Section~\ref{sec:inva}.\\

The simplified CBC method used in this paper considers the fundamental control component as the excitation applied to the system. The magnitude of the non-fundamental components of the control signal $u(t)$ is in general a measure of the controller invasiveness and hence of the accuracy of the experimental results --- a larger control signal is more likely to affect the position of the periodic responses. Considering the periodic nature of dynamics studied here, $u(t)$ can be decomposed as in Eq.~\eqref{eq:ctrl_sig_th}. This definition is convenient because the root problem solved by the CBC algorithm comprises as many equations as unknowns. However, in practice, the control signal will also include non-harmonic contributions coming, for instance, from noise. As such, the measured control signal takes the following general form
\begin{equation}
u(t) = \underbrace{A^u_1 \cos(\omega t)+B^u_1 \sin(\omega t)}_{\text{fundamental }=\text{ excitation}} + \underbrace{\frac{A^u_0}{2} + \sum^{m}_{j=2} A_j^u \cos (j \omega t) + B_j^u \sin (j \omega t) }_{\text{higher harmonics --- HH}} + \text{NH},
\label{eq:ctrl_sig_exp}
\end{equation}
where NH represents the non-harmonic components of the signal and is defined as the remainder of the decomposition of $u(t)$ into the first $m$ Fourier modes. As such, the two terms that contribute to the invasiveness of the method are the higher-harmonics (HH) and the non-harmonic (NH) components of the signal.\\

\begin{figure}[tbp]
\centering
\begin{tabular*}{0.95\textwidth}{@{\extracolsep{\fill}} c c}
\subfloat[]{\label{inv1}\includegraphics[width=0.456\textwidth]{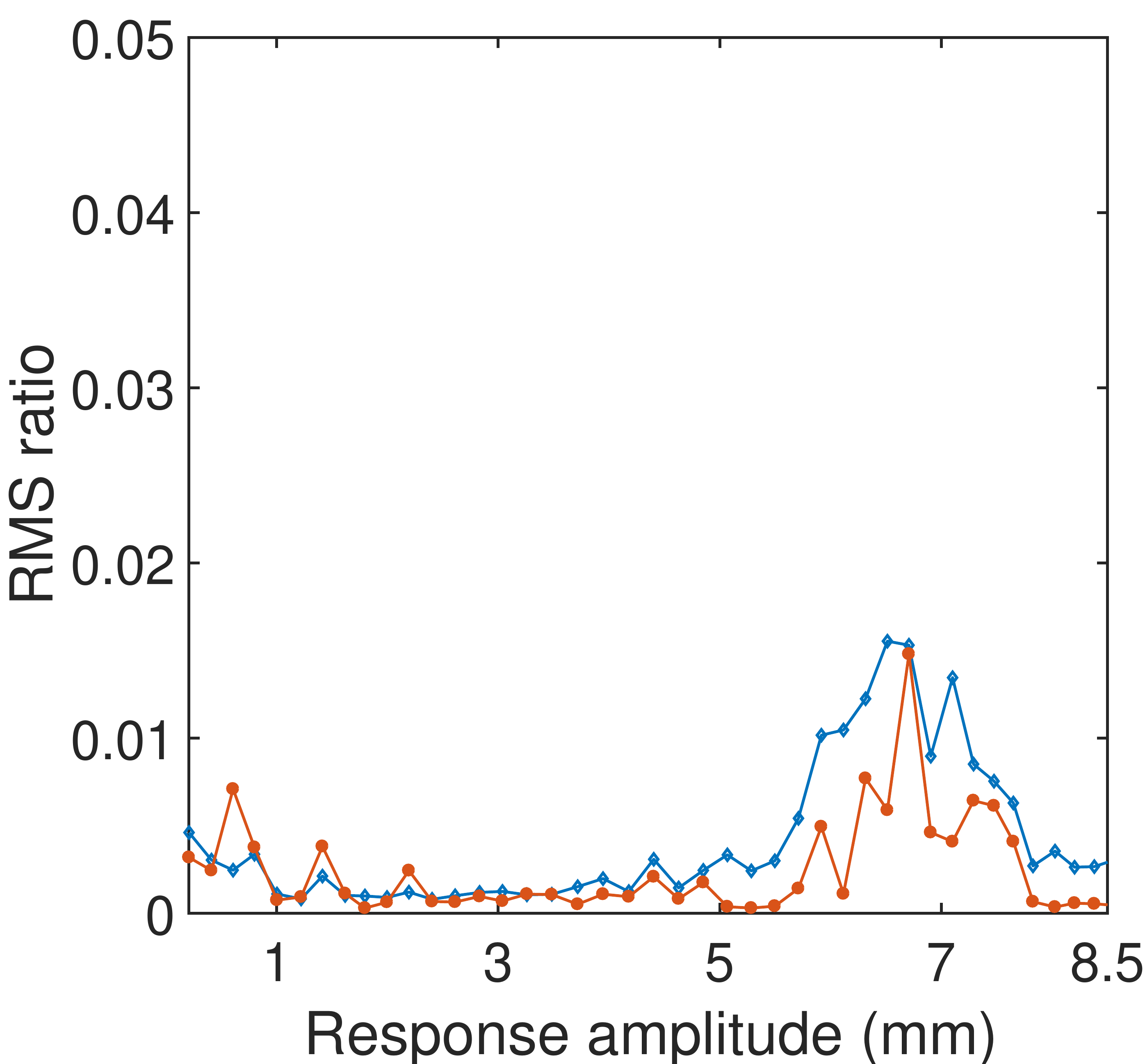}} &
\subfloat[]{\label{inv2}\includegraphics[width=0.48\textwidth]{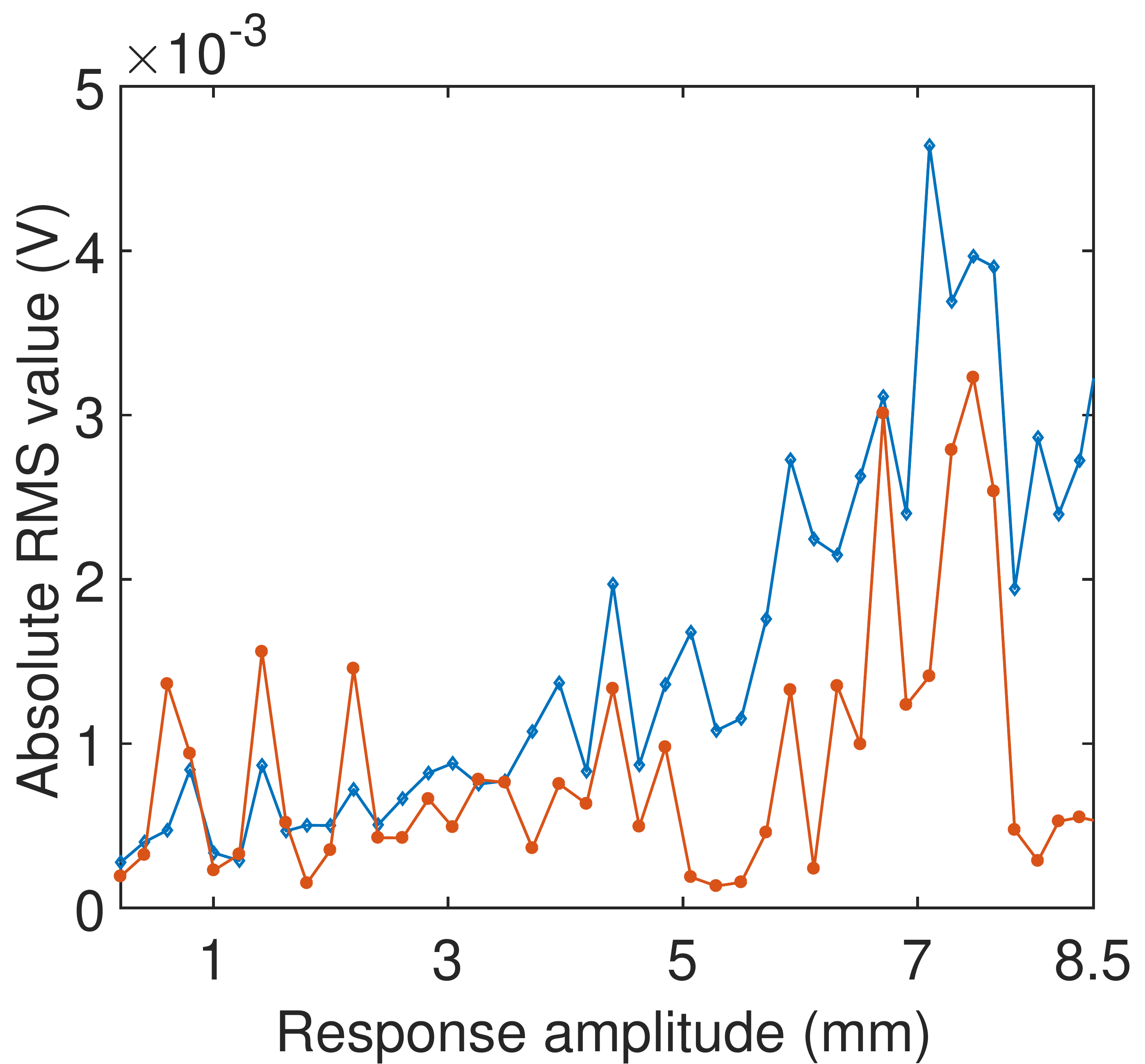}} \\
\end{tabular*}
\caption{Verification of the invasiveness of the control signal for one of the S-curves presented in Figure~\ref{fig:Scurve_repeat}. (a) RMS value of the higher-harmonic component (\textcolor{orange}{$\boldsymbol{-\bullet-}$}) and the non-harmonic component (\textcolor{blue}{$\boldsymbol{-\diamond-}$}) of the control signal relative to its fundamental component. (b) Absolute RMS value of the higher-harmonic component (\textcolor{orange}{$\boldsymbol{-\bullet-}$}) and the non-harmonic component (\textcolor{blue}{$\boldsymbol{-\diamond-}$}) of the control signal.}
\label{fig:inva}
\end{figure}

The relative importance of the HH (\textcolor{orange}{$\boldsymbol{-\bullet-}$}) and NH (\textcolor{blue}{$\boldsymbol{-\diamond-}$}) content of the control signal with respect to the fundamental excitation is analysed in Figure~\ref{fig:inva}(a) for one of the S-curves presented in Figure~\ref{fig:Scurve_repeat}. For each data point, the steady-state control signal was decomposed in its fundamental, higher-harmonic and non-harmonic components according to Eq~\eqref{eq:ctrl_sig_exp}. The root-mean-square (RMS) ratio was then used to compare the magnitude of the different signals.\\

The magnitude of the HH and NH components is mostly well below 1\%, which indicates that the total invasive contribution of the control signal, i.e. HH+NH, is small compared to the applied excitation. Furthermore, the amplitude of the HH content is usually found to be smaller than the amplitude of the NH content. This shows (indirectly) that the Picard iteration scheme used in the CBC algorithm of Section~\ref{sec:intro_cbc} converged for all the data points in the range of amplitudes considered. The figure also shows that to achieve a further reduction of the controller invasiveness the NH content must be addressed (in addition to the HH content). However, this is not possible using the algorithm of Section~\ref{sec:intro_cbc}.\\

For response amplitudes between approximatively 6mm and 7.5mm, larger HH and NH content is observed. The relative measure used in the figure is one of the factors that contribute to the appearance of this bulge. Indeed, this region corresponds to the resonance of the system where the response amplitude is large compared to the fundamental excitation. The HH and NH contents will therefore appear relatively larger in this region than in other non-resonant regions. The absolute RMS value of the HH content is also observed to increase around the resonance region (Figure~\ref{fig:inva}(b)). This is thought to be caused by key higher-harmonics in the force applied to the structure. This aspect is further discussed in this section. This could have been reduced by adjusting the tolerance $\delta$. However, this was deemed unnecessary given the small values of the HH and NH signals ($\leq 5 \times 10{-3}$ V). Figure~\ref{fig:inva}(b) also shows that the NH content tends to increase with the oscillation amplitude. A spectral analysis of the NH time signal reveals that this increase is due to the emergence of additional harmonics (higher than the 7 already considered) and to an overall larger level of noise, which could be caused by a degradation of the performance of the controller in this region or large amplitude vibrations causing other behaviour which is sensed and feed through the controller.\\

\begin{figure}[tbp]
\centering
\includegraphics[width=0.45\textwidth]{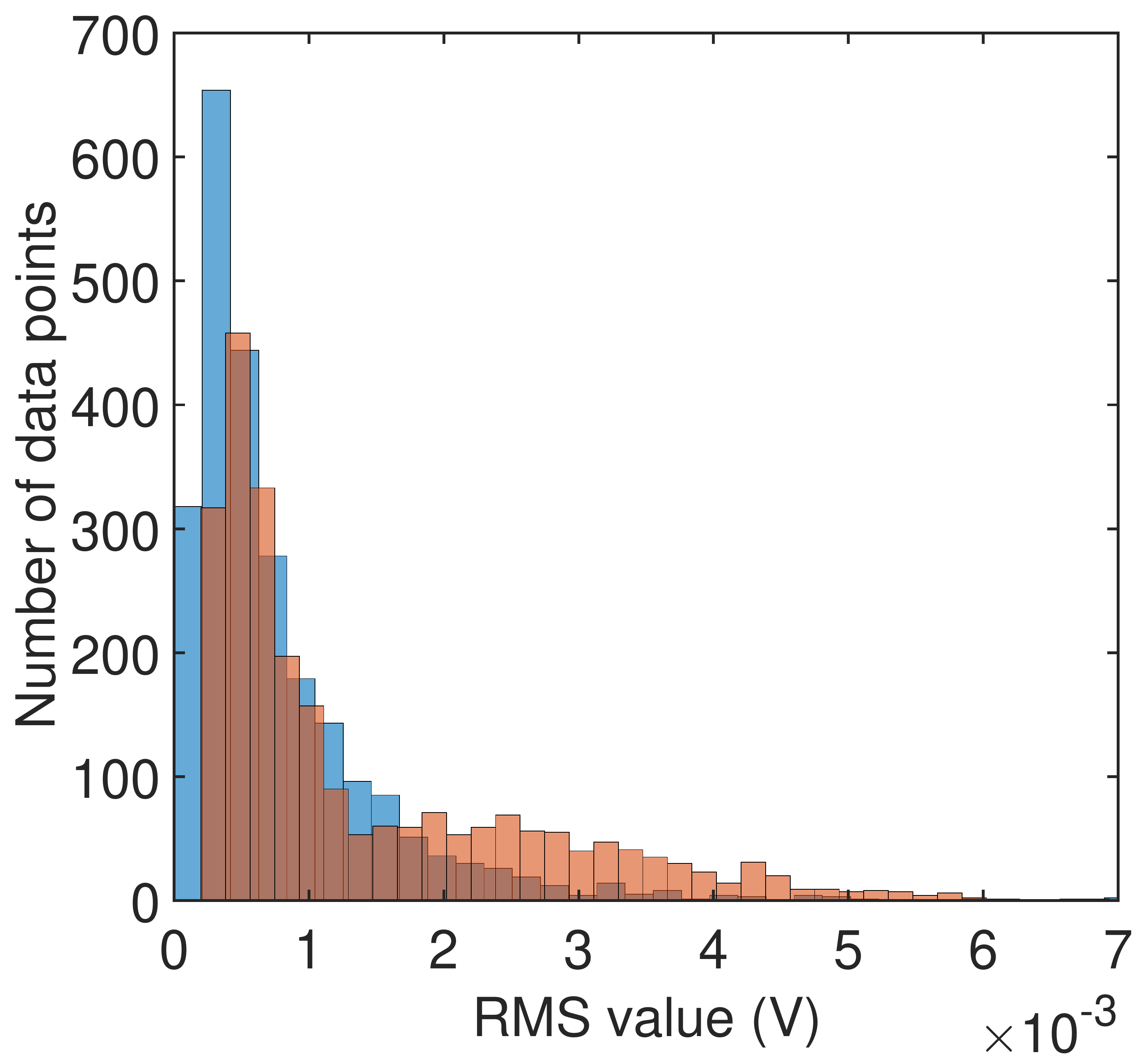}
\caption{Verification of the invasiveness of the control signal. Histogram of the RMS value of the higher-harmonic component (\textcolor{orange}{$\blacksquare$}) and the non-harmonic component (\textcolor{dblue}{$\blacksquare$}) of the control signal for all the data points collected.}
\label{fig:inva2}
\end{figure}

Following the same decomposition of the control signal, Figure~\ref{fig:inva2} presents an histogram of the absolute RMS values of the HH (\textcolor{orange}{$\blacksquare$}) and NH (\textcolor{dblue}{$\blacksquare$}) components for all the data points collected experimentally. The vast majority of the data points presents RMS values that are smaller that 5$\times$10$^{-3}$ Volts, which further demonstrates the small magnitude of the invasive control effort. However, contrary to Figure~\ref{fig:inva}, it appears that a larger proportion of data points may have a HH content that is greater than the NH content. Although this could probably be improved by performing additional Picard iterations, this was not deemed necessary given the small RMS values at issue. Twelve data points present RMS values much larger than 7$\times$10$^{-3}$ Volts for either the HH or the NH content. These data points were considered as inaccurate and discarded before analysing the results.\\

\subsection{Comparison between open- and closed-loop results}\label{sec:inva}
To further validate the experimental results and demonstrate that the observed dynamics corresponds to the dynamics of the underlying uncontrolled experiment, some of the data points obtained using CBC are directly compared to data points collected in open-loop conditions. To this end, each data point considered is first reached using feedback control. An excitation identical to the one given by the fundamental control components $(A_1^u,B_1^u)$ is then applied to the experiment and the control signal is simultaneously switched off. In this way, the controller is no longer active but the fundamental input to the experiment remains unchanged. After waiting 50 seconds for any transient dynamics to develop and then settle, the open-loop response of the system was recorded. This procedure is applied to all the data points of the 1.5N frequency response curve presented in Figure~\ref{fig:nlfr_slices}(c).\\

\begin{figure}[tbp]
\centering
\begin{tabular*}{1.\textwidth}{@{\extracolsep{\fill}} c c}
\subfloat[]{\label{stab1}\includegraphics[width=0.45\textwidth]{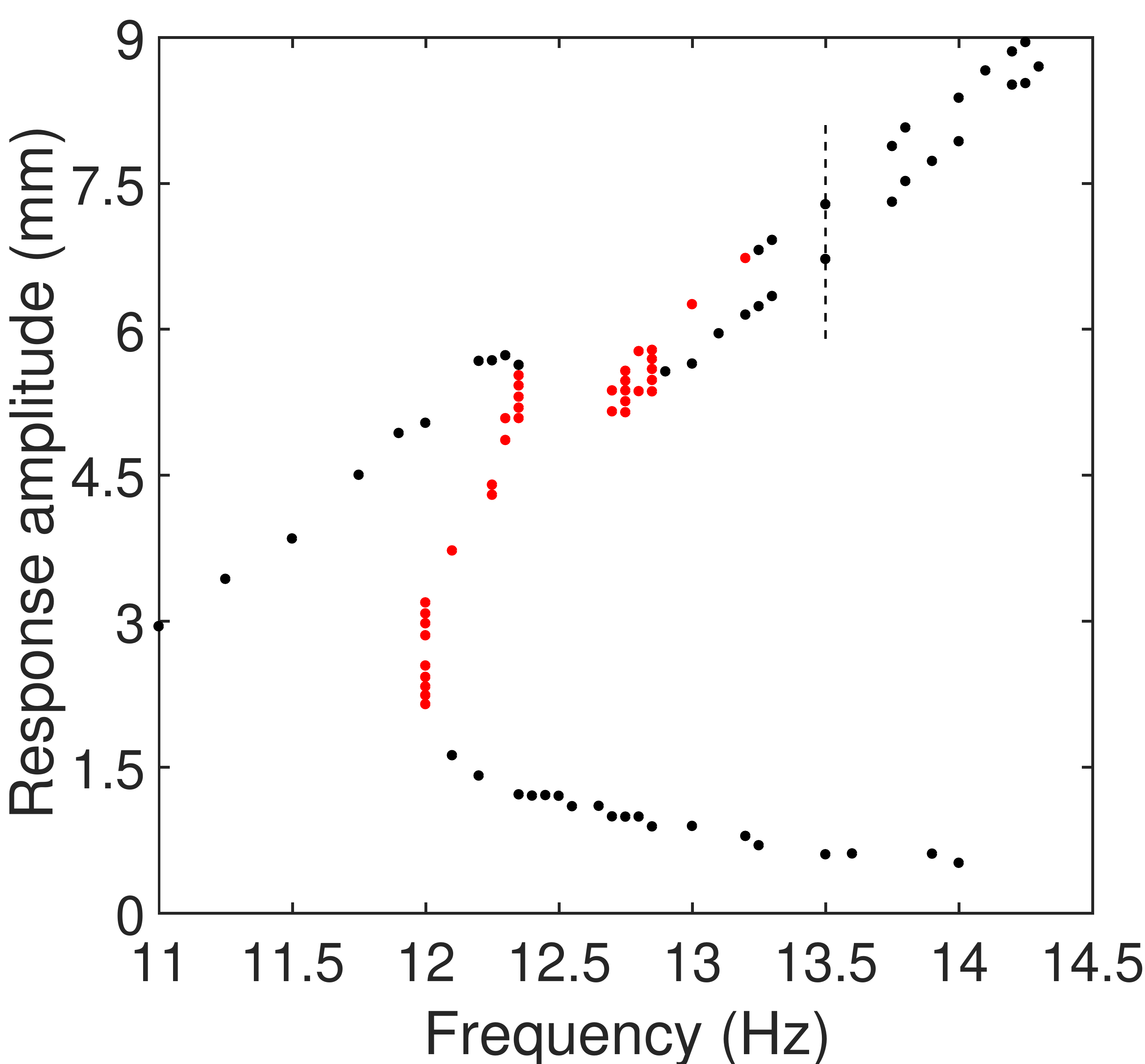}} &
\subfloat[]{\label{stab4}\includegraphics[width=0.49\textwidth]{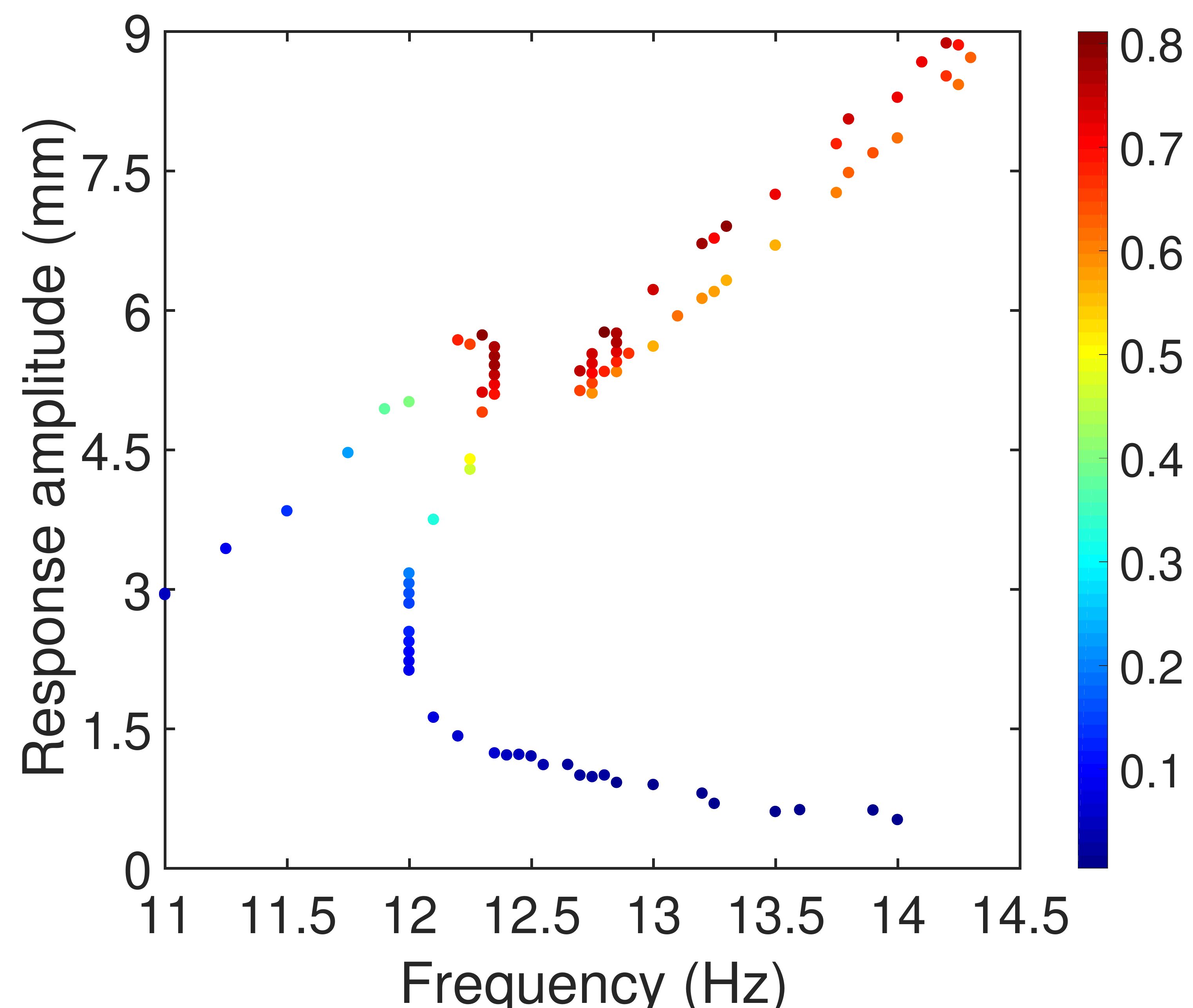}}
\end{tabular*}
\caption{Nonlinear frequency response of the system at 1.5 N (as in Figure~\ref{fig:nlfr_slices}(d)). (a) Closed-loop data points with colouring indicating closeness to open-loop measurements --- black indicates an absolute difference smaller than 0.2~mm and a relative difference smaller than 1\%. Otherwise, points are in red. (b) Total amplitude of the higher harmonics in the applied force compared to the fundamental.}
\label{fig:stab_1p5N}
\end{figure}

The result of the comparison between closed- and open-loop measurements is presented in Figure~\ref{fig:stab_1p5N}(a) where closed-loop data points are coloured according to the difference in response amplitude observed between the two tests. Points that present an absolute difference smaller than 0.2mm and a relative difference smaller than 1\% are in black. Those points typically correspond to stable responses of the system and further demonstrate the validity of the results obtained with CBC. The points that do not satisfy the above criterion are coloured in red. When the controller is turned off at a red point, the dynamics of the system evolves to coexisting stable attractors. Responses on the main resonance peak were all found to reach the other periodic responses identified by the CBC method. Responses on the isola evolved towards low-amplitude periodic responses or quasi-periodic oscillations (as observed in Section~\ref{sec:step-sine}). All the red points are thought to correspond to unstable periodic responses of the underlying uncontrolled experiment. They cannot be observed experimentally without control. As such these points cannot be used to assess CBC results.\\

An important observation can be made when looking at the stability of the isola; the responses located on the lower branch are mostly stable. This is surprising given that theoretical investigations usually show that responses found there are unstable. Take, for example, the results obtained in~\cite{Shaw16} where the same system as the one studied here was numerically investigated. Here, open-loop tests clearly show that these responses are stable. The most significant difference between the experiment and the theoretical results presented in~\cite{Shaw16} is the presence of higher harmonics in the force applied to the structure. These higher harmonics can be important as shown in Figure~\ref{fig:stab_1p5N}(b) where their total contribution to the excitation can reach up to 80\% of the excitation at the fundamental component. However, these higher harmonics appears to have a limited influence on the higher harmonics present in the response of the system as shown in Figure~\ref{fig:h3_nlfr} for the third harmonic.\\

\begin{figure}[htbp]
\centering
\begin{tabular*}{1.\textwidth}{@{\extracolsep{\fill}} c c}
\subfloat[]{\label{stab1}\includegraphics[width=0.45\textwidth]{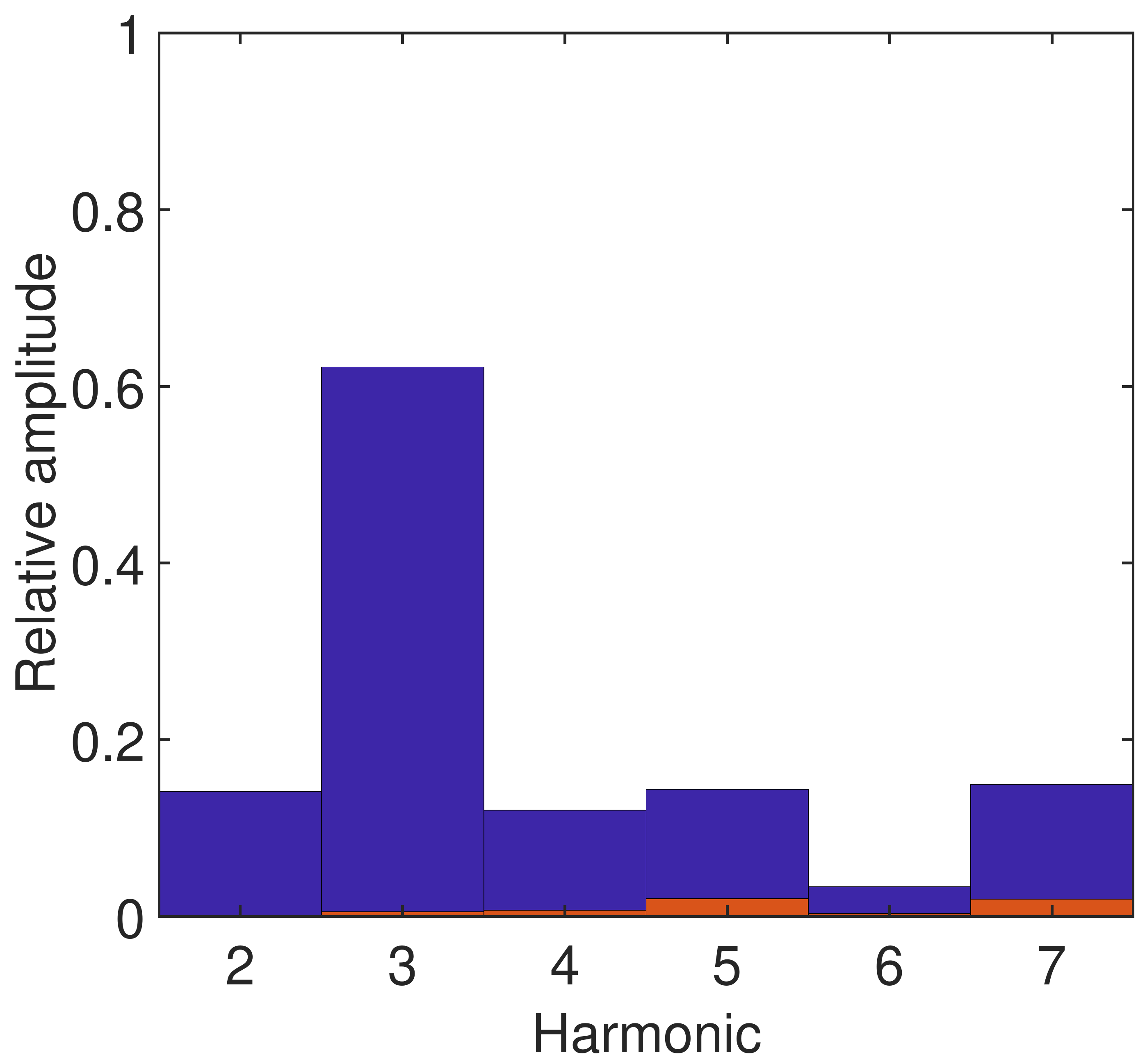}} &
\subfloat[]{\label{stab4}\includegraphics[width=0.45\textwidth]{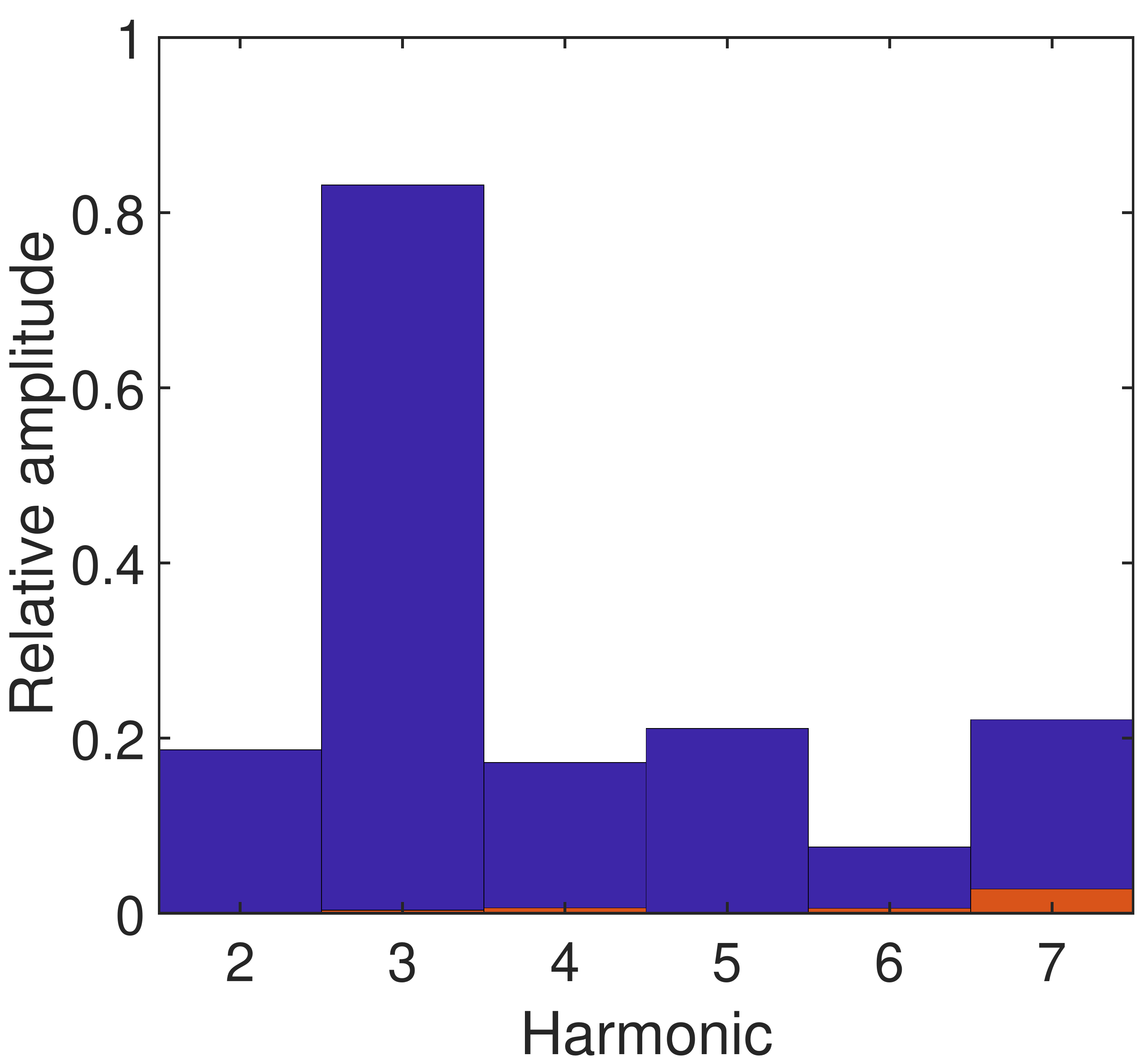}}\\
\subfloat[]{\label{stab1}\includegraphics[width=0.45\textwidth]{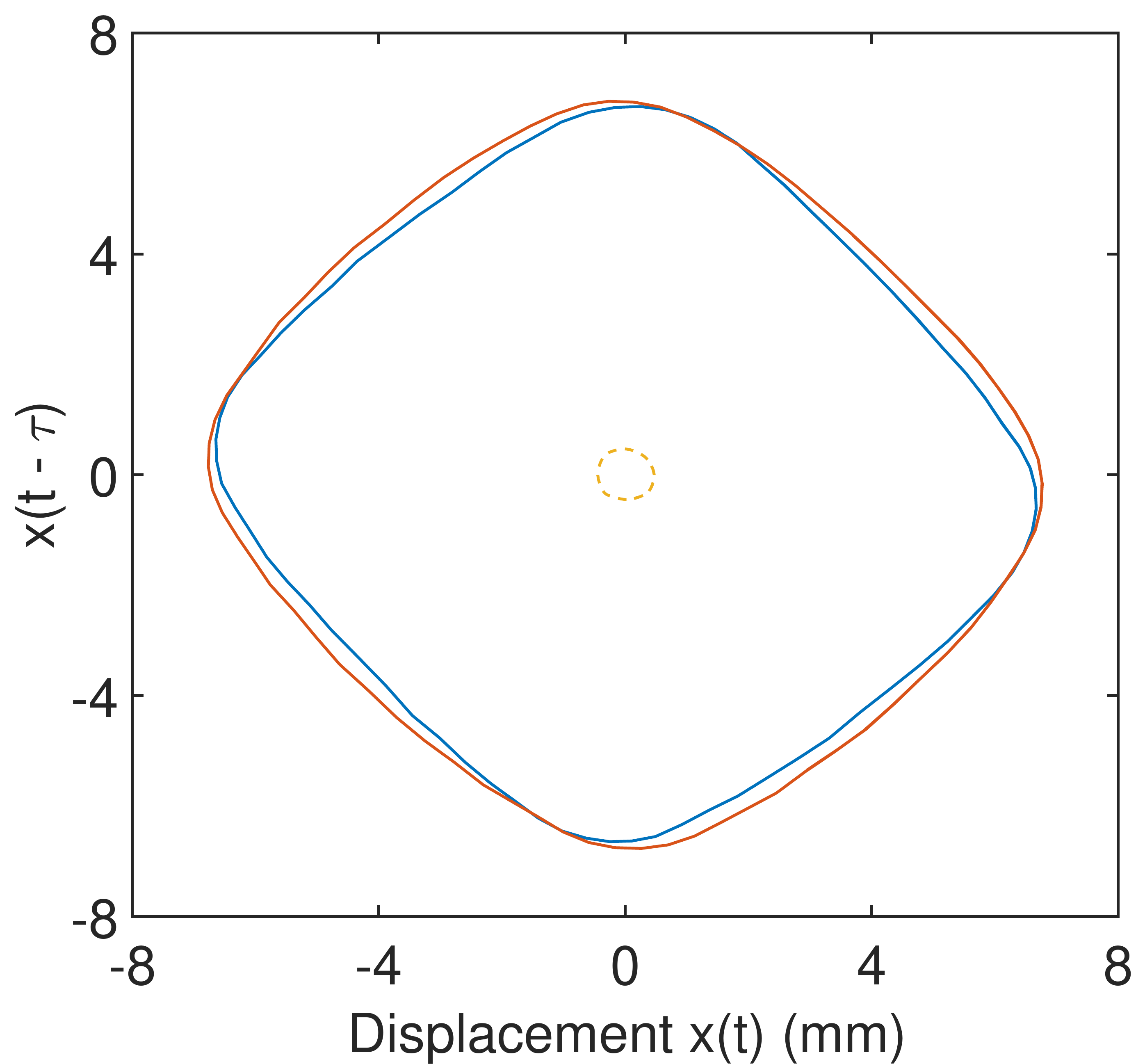}} &
\subfloat[]{\label{stab4}\includegraphics[width=0.45\textwidth]{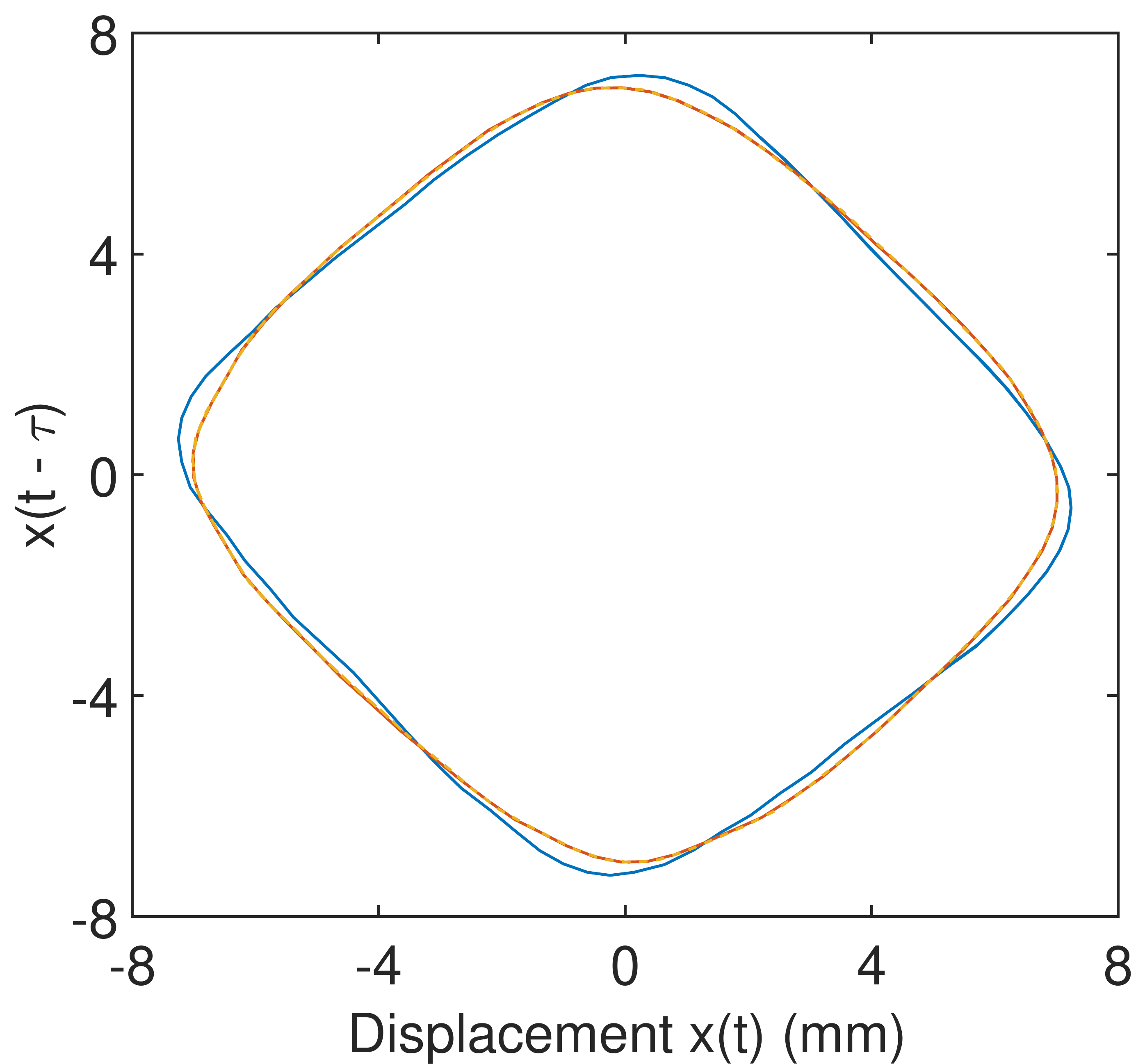}}
\end{tabular*}
\caption{Effect of the higher harmonics in the applied force on the dynamic behaviour of the system for the lower (a, c) and the upper (b, d) periodic solutions highlighted in Figure~\ref{fig:stab_1p5N}(a). (a, b) Relative importance of the higher harmonics compared to the fundamental component of the force applied to the system. (c, d) Phase portraits of the steady-state periodic orbits. Closed-loop experimental conditions before (\textcolor{dblue}{$\blacksquare$}) and after (\textcolor{orange}{$\blacksquare$}) applying the harmonic compensation procedure. Open-loop experimental conditions (\textcolor{dyellow}{$\blacksquare$}).}
\label{fig:stab_1p5N_force}
\end{figure}

The existence of these higher-harmonics in the force is well known in the literature~\cite{Claeys14,Chen16} but they are typically ignored, and only rarely are their effects on the dynamic behaviour of the system studied experimentally. To observe experimentally their effect on the stability of the response, the force applied to the structure was corrected to be purely harmonic. This was achieved in CBC using an iterative harmonic compensation procedure which considers the introduction of higher harmonics in the voltage input of the shaker to cancel out the higher harmonics present in the force~\cite{Renson17}. This procedure was applied to the two data points highlighted by a dashed line in Figure~\ref{fig:stab_1p5N}(a). The harmonic content of the force relative to the fundamental component is shown before applying the correction procedure in blue (\textcolor{dblue}{$\blacksquare$}) in Figures~\ref{fig:stab_1p5N_force}(a) and~(b) for the low- and high-amplitude points, respectively. After correction, a significant reduction of the higher harmonic content in the force signal is achieved (\textcolor{orange}{$\blacksquare$}) though there remains a significant higher harmonic content in the response. Note that, whilst the higher harmonics are removed, the amplitude of the fundamental component of the force was kept equal to 1.5N such that both points still belong to the same nonlinear frequency response curve.\\

Following the correction of the excitation, the controller was turned off using the same procedure as described earlier. This time the low-amplitude response was observed to be unstable whereas the high-amplitude response was found to be stable. This is illustrated in Figures~\ref{fig:stab_1p5N_force}(c)~and~(d) where the phase portrait of the steady-state periodic orbits obtained in closed-loop after correction of the higher harmonics (\textcolor{dorange}{$\boldsymbol{-}$}) is compared to the open-loop periodic orbit (\textcolor{dyellow}{$\boldsymbol{-}$}) for the low- and high-amplitude points, respectively. The presence of the higher harmonics in the force is also found to have a small effect on the shape of the orbit as shown by the portraits obtained before (\textcolor{dblue}{$\boldsymbol{-}$}) and after (\textcolor{dorange}{$\boldsymbol{-}$})  applying the correction procedure.\\

The objective of the above investigation was to validate the results obtained using CBC, rather than analyse the stability properties of the system. A similar procedure where the controller is momentarily turn off was used in~\cite{Bureau14} to analyse the stability of the periodic responses of an impacting beam. This approach presents the risk that the controller may not be able to cope with the generated transient dynamics. The experiment could then be damaged by reaching undesirable, high-amplitude responses. A method of working out the original stability properties of the underlying uncontrolled system without turning off the controller was presented in~\cite{Barton16}. However, regardless of the chosen method, the present results suggest that a precise understanding of the stability of the responses measured using CBC may require careful control of the force applied to the system. This can be achieved using, for instance, the harmonic compensation procedure considered here but we note that applying this method was time consuming. We also note that this harmonic compensation may not be necessary or even desirable if the experimental results are to be used for the development and validation of a mathematical model of the structure as higher-harmonics can improve the accuracy of parameter estimations~\cite{Chen16,Chen17}.\\

\FloatBarrier
\section{Conclusions}\label{sec:conclusion}
Control-based continuation is a general and systematic method for exploring the nonlinear dynamics of physical experiments. To illustrate the power of this method, it was applied to a nonlinear beam structure that exhibits a strong 3:1 modal coupling between its first two bending modes. Control-based continuation was able to extract a range of dynamical features, including an isola, directly from the experiment without recourse to model fitting or other indirect data-processing methods. The invasiveness of the controller used was analysed and experimental results were validated with open-loop measurements. Good agreement between open- and closed-loop results was achieved, though it was found that the presence of higher-harmonics in the force applied to the structure can affect the stability properties of the responses.\\
 
Previously, control-based continuation has only been applied to (essentially) single-degree-of-freedom experiments. This paper has experimentally demonstrated that this method is applicable to multi-degree-of-freedom systems exhibiting complex multi-mode dynamics. We showed that the feedback-control principles and path-following techniques previously used on single-degree-of-freedom systems can equally be applied to multi-degree-of-freedom systems. Furthermore, the complexity of control-based continuation appears to scale with the dimension of the instability rather than the dimension of the system.\\

A low-level broadband excitation was initially applied to the experiment to obtain the requisite information for controller design. The controller was deliberately kept simple to show that control-based continuation may not in general require complex control strategies. Specifically, the feedback law used here was linear and designed using standard techniques readily available in the literature. Subsequently, the physical experiment was treated as a “black box” that is probed using control-based continuation.

\section*{Data Statement}
Experimental data collected in this study are available at [\textit{DOI to be inserted at proofing}].

\section*{Acknowledgements}
L.R. has received funding from the Royal Academy of Engineering, fellowship RF1516/15/11. D.A.W.B. is funded by the EPSRC grant EP/K032738/1 and S.A.N. by the EPSRC fellowship EP/K005375/1. We gratefully acknowledge the financial support of the Royal Academy of Engineering and the EPSRC.

\bibliographystyle{unsrt}
\bibliography{mybib}

\end{document}